\documentclass[11pt,letter]{article}
\usepackage{jheppub}
\usepackage{slashed}
\usepackage{physics}
\usepackage{amsmath}
\usepackage{pgfplots}
\usepackage{bbm}
\usepackage{nicematrix}
\usepackage{tensor}
\usepackage{multirow}
\usepackage{booktabs}
\makeatletter
\def\@fpheader{\relax}
\makeatother
\usepackage{graphicx}
\usepackage{wrapfig}
\usepackage{empheq}
\usepackage{ulem}
\usepackage{wrapfig}
\usepackage[czech,english]{babel}
\usepackage{graphicx}
\usepackage{amsmath,amsfonts,amssymb}
\usepackage{url}
\usepackage{mathtools}
\usepackage{float}
\usepackage{euscript}
\usepackage[makeroom]{cancel}
\usepackage{leftidx}
\usepackage{xcolor}
\numberwithin{equation}{section}
\usepackage{adjustbox}
\usepackage{appendix}
\usepackage{subcaption}
\usepackage{tikz}
\tikzstyle{process} = [rectangle, minimum width=3cm, minimum height=1cm, text centered, draw=black, fill=orange!30]
\tikzstyle{arrow} = [thick,->,>=stealth]

\usepackage{comment}

\preprint{LCTP-24-14}

\title{Supergravity Spectrum of AdS$_5$ Black Holes}

\author[a]{Nizar Ezroura}

\author[a]{Finn Larsen}

\emailAdd{nezroura@umich.edu}

\emailAdd{larsenf@umich.edu}

\affiliation[a]{Leinweber Center for Theoretical Physics, University of Michigan, Ann Arbor, MI 48109, U.S.A.}

\abstract{

We embed Kerr-Newman-AdS black holes into $\mathcal{N}=8$ gauged supergravity and study quadratic fluctuations 
around the black hole backgrounds of all fields in the larger theory. The equations of motion of the perturbations are partially diagonalized by the group theory of broken symmetry. Nearly all fields in theory have non-minimal couplings, so their equations of motion are not merely massive Klein-Gordon equations with minimal coupling to background gauge fields, and their analogues for fields with spin. In the special case of extremal black holes we identify specific modes of instability, some of which touch supersymmetric locus. For example, we identify scalar fields in supergravity that condense in the near horizon region and transition the black hole into a superconducting phase. We also identify supergravity modes that are susceptible to superradiant instability. 
}

\keywords{}

\arxivnumber{}

\begin{document}

\maketitle


\section{Introduction}

An extremal black hole is the lightest regular solution to the field equations of general relativity in a superselection sector defined by a set of conserved charges. That the state has the lowest energy suggests an interpretation as the ground state of the system which, 
because of the huge back hole entropy, should be exceptionally stable. On the other hand, intrinsic repulsion between charged constituents and/or the centrifugal force due to rotation arguably renders extremal black holes on the verge of instability. The balance between these intuitions is far from obvious and depends on the particulars of boundary conditions, spacetime dimension, and the matter content of the theory that the black hole solves. Some discussions are \cite{Arkani-Hamed:2006emk,durkee2011perturbations, horowitz2023almost}. 

Supergravity and its string theory progenitors can illuminate these questions. In this article we consider an important generic class of 
AdS$_5$ black holes that has a single angular momentum and a single electric charge  \cite{chong2005general}. We refer to these black holes as Kerr-Newman black holes, in analogy with their familiar relatives in four dimensions. We show how they can be interpreted as solutions in gauged ${\cal N}=8$ supergravity which, in turn, is holographically dual to ${\cal N}=4$ SYM. In this general setting, we show that quadratic fluctuations of
{\it all} $256$ degrees of freedom decouple in ``blocks": they are independent except that, in general, the background black hole couples them in {\it pairs}. The resulting equations of motion are very explicit and manageable, also for particles with spin, without any assumptions on the parameters of the background black hole.

Importantly, the field equations we find are not ``minimal", in the sense that they do not reduce to the Klein-Gordon equation for a massive scalar field in a gravitational background, or its analogues for particles with spin. Most fields have minimal couplings to the background gauge field $A_\mu$, and many also have non-minimal couplings directly to the background field strength $F_{\mu\nu}$. Therefore, fluctuations of scalar and vector fields can couple in the black hole environment. The non-minimal couplings can rarely be neglected, except in the AdS-Kerr background, when the field strength vanishes. They are important even for the qualitative behavior. For example, they are needed to establish that supersymmetric black holes are stable to perturbations at linear order. 

The AdS/CFT correspondence relates the fall-off of supergravity fields far from the black hole to the conformal weights of operators in the dual ${\cal N}=4$ SYM. Our result gives the field equations that follow such perturbations all the way to the horizon region. Although our field equations apply broadly, the main application we have in mind is to extremal black holes in AdS$_5$, whether supersymmetric or not. In this context, our work shows how fields in ${\cal N}=8$ supergravity reorganize from their asymptotic behavior, identifiable in terms of familiar operators in ${\cal N}=4$ SYM, into fields in the near horizon AdS$_2$-region. These results detail the onset of instability. 

In AdS$_5$ supergravity with ${\cal N}=8$ supersymmetry there are ${\bf 20}'$ modes with conformal weight $\Delta=2$ \cite{Gunaydin:1985cu, Kim_Romans_van_Nieuwenhuizen_1985, ferrara1998n}. Equivalently, these fields have a mass that is at the Breitenlohner-Freedman bound \cite{breitenlohner1982positive}. That means they are marginal in that, if they had been any lighter, their modes would increase exponentially. The black hole background deforms the couplings of these fields. It turns out that $8$ of the ${\bf 20}'$ scalar fields are neutral with respect to the electric field of the black hole, but couple to fluctuations of $8$ partner vector fields that vanish in the black hole background, but are an unavoidable part of the supergravity spectrum. The remaining $12$ scalars in the ${\bf 20}'$ couple to the field strength of the electromagnetic field radiating from the Kerr-Newman black hole. They do not only have a conventional minimal gauge coupling, implemented by the gauge covariant derivative $\nabla_\mu + i e A_\mu$, but also a Pauli coupling directly to the gauge invariant field strength $F_{\mu\nu}$. The latter are less familiar because they are irrelevant according to the standard counting of derivatives. However, the minimal and Pauli couplings are comparable in the vicinity of a Kerr-Newman black hole. This is one of the situations we address. 

A scalar field with charge $e$ that couples minimally to an electric field receives a contribution to its effective mass $m^2 = e^2 g^{tt} A_t^2$ that is negative (in our signature $-+++$). Depending on other contributions, it may be energetically favored for the scalar to develop an expectation value and, where that happens, a superconductor (or more precisely a superfluid) has formed. For example, magnetic flux is expelled from the region, and the electric resistance vanishes. In the right circumstances, this symmetry breaking pattern is realized near a black hole. This scenario was studied in many ``toy" models motivated by the AdS/CFT correspondence \cite{Zaanen_Schalm_Sun_Liu_2015, Gubser_2008, nishioka2010holographic,hartnoll2008building,hartnoll2008holographic}, and in some top-down models \cite{Gubser:2009qm,Gauntlett:2009dn}. Our work adds examples from the theory with maximal supersymmetry.  

Nontrivial matter fields outside the black hole horizon, such as the scalar condensates responsible for superfluidity are examples of black hole ``hair" \cite{liu2007new, basu2010small, dias2010scalar, brihaye2012scalar, dias2012hairy, markeviciute2018rotating, markevivciute2018evidence, choi2024}. In other words, 
this is structure of the black hole solution that is not determined by the conserved charges $M, J, Q$. That this situation is even possible is contrary to conventional wisdom, 
or lore, but it does not violate any firm principles. Moreover, the amount of hair involved is tiny compared to the vast entropy associated with the black hole interior, so it remains negligible for some purposes. 

There is a huge literature on black hole instability, including much recent work, so it is not realistic to do justice to all aspects of the subject. To position our results in the broader context, in the following we highlight a few important developments.  

\bigskip
\begin{itemize}
\item 
{\bf Negative Specific Heat}

The standard Schwarzchild black hole in four asymptotically flat spacetime dimensions is {\it unstable to emission of Hawking radiation}. Therefore, the true equilibrium is not a black hole, it is the final gas of particles. However, there are essential caveats. The decay takes extraordinarily long time, and the final gas will never equilibrate in infinite space. A more physical setup would enclose the Schwarzchild black hole in a huge, but finite, box \cite{hawking1976black, gibbons1978black}. Then the ground state configuration would be a black hole in equilibrium with a gas at the Hawking temperature. These distinctions are precise in AdS, where ``small" and ``large" black holes correspond to the unstable/stable black hole without/with enclosure in a box \cite{hawking1983thermodynamics,Witten:1998zw}. 

The instabilities that motivate our work, and some aspects of recent research \cite{Horowitz:2023xyl,Kim:2023sig}, are {\it not} at all like the quantum instability due to emission of Hawking radiation. They are much more serious.

\bigskip
\item
{\bf Superradiance} 

Adding rotation, the standard Kerr black hole in four asymptotically flat spacetime dimensions is also unstable to emission of Hawking radiation. However, in addition, it suffers {\it classical instability to superradiance}. Radiation scattering off the black hole can be reflected with larger amplitude, and so extract energy \cite{penrose1971extraction}. This mechanism is classical, and it is not cured by enclosure in a large box. Quite the contrary, radiation making it to the boundary of the box is reflected back into the black hole, where it stimulates even more radiation. The resulting runaway behavior has been dubbed the black hole bomb \cite{press1972floating, zel1972amplification, starobinskii1973amplification, vilenkin1978exponential, cardoso2004black, green2016superradiant}. It has been speculated that this mechanism may be relevant for the behavior of astrophysical black holes \cite{Aretakis:2011gz,Aretakis:2023ast}. 

Superradiance {\it is} relevant for recent developments. The modern way to introduce a box is to study black holes with asymptotically AdS boundary conditions. In a statistical ensemble with boundary conditions corresponding to specification of rotational velocity $\Omega$ and electric potential $\Phi$, black holes are unstable to superradiance if, in the units we used in \cite{larsen2020ads5}, either $\Omega>1$ or $\Phi>3$. {\it All} non-supersymmetric extremal black holes in ${\cal N}=8$ supergravity either have $\Omega>1$ or $\Phi> 3$ and are thermodynamically unstable. The two forms of instability meet at the BPS locus where $\Omega=1$ and $\Phi=3$. 
 
\item
The {\bf  Weak Gravity Conjecture} (WGC) was originally motivated, at least in part, by the presumption that, in a consistent theory of quantum gravity, all extremal black holes must have suitable quantum states to which they can decay \cite{Arkani-Hamed:2006emk}. It would be circular to claim that this establishes the instability of extremal black holes. However, the failure of significant efforts to find counter-examples to the WGC in (string) theories of quantum gravity that exist in the UV amounts to persuasive evidence in favor of this conclusion \cite{Harlow:2022ich}.

The WGC complements the statement that extremal black holes always have $\Omega\geq 1$ or $\Phi\geq 3$ by claiming that suitable decay channels exist. Our work contributes to this program by identifying unstable supergravity modes explicitly.  

\item
{\bf Quantum Corrections} 

It was recognized a long time ago that the thermal description of black holes is not viable at {\it very} low temperature, or else individual thermal fluctuations would require the entire available energy \cite{Preskill:1991tb, Maldacena:1998uz}. In the last few years it was understood that the approach to zero temperature is described by the Schwarzian theory, or one of its extensions, interpreted as an effective quantum theory \cite{Maldacena:2016upp}.
According to the exact solution of this low energy theory, supersymmetric black holes feature a huge ground state degeneracy protected by a finite gap, but extremal black holes that do not preserve supersymmetry have no ground states at all
\cite{Stanford:2017thb, Heydeman:2020hhw, Iliesiu:2020qvm,Boruch:2022tno, turiaci2023new}. These results suggest 
non-supersymmetric extremal black holes would suffer from a {\it quantum} instability and so they cannot exist. 

The analysis in low energy effective field theory does not by itself identify the nature of the black hole instability, nor the true ground state of the quantum system. Our analysis of the low energy spectrum in explicit theories contributes towards this goal.

\end{itemize}

This article is organized as follows. 

In section \ref{sec:N=8thy}, we briefly review the basic equations of ${\cal N}=8$ gauged supergravity in five dimensions. This serves to introduce our notation. 

In section \ref{sec:KNAbknd} we show how any solution to standard Einstein-Maxwell-AdS theory in five dimension can be interpreted as a solution to 
${\cal N}=8$ supergravity. This involves specifying a linear combination of the $15$ $SU(4)_R$ vector fields in ${\cal N}=8$ supergravity
that is ``the" gauge field in Einstein-Maxwell-AdS theory, and so non-zero in the black hole solution. We show that, whatever the configuration of this preferred linear combination of gauge fields, the equations of motion are satisfied when all other gauge fields vanish, and all scalar fields are constant. 

In section \ref{sec:fluc} we expand the original ${\cal N}=8$ theory  to quadratic order around any background that solves the Einstein-Maxwell-AdS theory. This gives the equations of motion for such fluctuations. Because the embedding preserves an $SU(1,2|3)$ subgroup of the  $PSU(2,2|4)$ global symmetry of the original ${\cal N}=8$ supergravity, the fluctuations are organized into ``blocks" that relate fields according to their representation of $SU(1,2|3)$. In practice, that means the fields of ${\cal N}=8$ supergravity couple in pairs, 
but never in larger groups. 

In section \ref{sec:nearhg} we study the extremal KNAdS black hole in its near horizon region. The geometry is fairly simple, it is just AdS$_2$ with $S^2\times S^1$ fibered over it. However, there is a nontrivial relation between the conserved black hole charges and the black hole solution, comprising both geometry and the gauge fields that supports it. We also consider generic fields in the near horizon background.  

In section \ref{sec:nearh} we solve the equations of motion of the bosonic ${\cal N}=8$ supergravity fields in the near horizon geometry. 
We decouple the differential equations and express the results in terms of effective masses in AdS$_2$ that are equivalent to conformal dimensions of operators in the dual (nearly) CFT$_1$. This gives regions of (in)stability, as measured by the BF-bound in AdS$_2$. As special cases, we highlight the families of extremal Kerr-AdS, BPS (supersymmetry), and Reissner-Nordstr\"om-AdS black holes. 

Our main technical results are summarized in sections \ref{subsec:fluceom} and \ref{subsec:fluceffmass}, especially the tables therein. Section \ref{sec:discussion} has final discussion and prospects for future research.

\section{$\mathcal{N}=8$ Gauged Supergravity: Field Content and Equations of Motion}
\label{sec:N=8thy}

In this section we introduce the field content of gauged and ungauged 
$\mathcal{N}=8$ supergravity in five dimensions. We present both the Lagrangian and the equations of motion of all fields. We stress transformation properties under symmetries that will be important later. 

\subsection{$\mathcal{N}=8$ Ungauged Supergravity}

We start with the field content of $\mathcal{N}=8$ ungauged supergravity
in five dimensions. It is organized according to representations of $\text{USp}(8)$, the global symmetry of the theory:
\begin{itemize}
\item $\mathbf{1}$ graviton field $g_{\mu\nu}$ (spin-2, 5 on-shell d.o.f.) , 
\item $\mathbf{8}$ gravitini $\psi_{\mu a}$ (spin$-\tfrac{3}{2}$, $8\times 4=24$ d.o.f.) , 
\item $\mathbf{27}$ vectors $A_{\mu ab}$ (spin-1, $27 \times 3=81$ d.o.f.) ,
\item $\mathbf{48}$ gaugini $\lambda_{abc}$ (spin-$\tfrac{1}{2}$, $48 \times 2 = 96$ d.o.f.) ,
\item $\mathbf{42}$ scalars with tangent vectors $P_{\mu abcd}$ (spin-0, 42 d.o.f.) .
\end{itemize}
Each of these fields field is fully antisymmetric and symplectic traceless in the $USp(8)$ vector indices $a, b, \ldots $. In total, there are $128$ bosonic and $128$ fermionic on-shell degrees of freedom. 

The full structure of the theory depends crucially on the nonlinearly realized duality group $\text{E}_{6(6)}$. The field strength $F_{\mu AB}=V_{AB} ^{ab}F_{\mu ab}$ transforms in the fundamental ${\bf 27}$ of $\text{E}_{6(6)}$, which is identified with the antisymmetric tensor of $USp(8)$ through the $27$-bein $V_{AB} ^{ab}$ that realizes the $\text{E}_{6(6)}/USp(8)$ coset. The adjoint ${\bf 78}$ of $\text{E}_{6(6)}$ branches to ${\bf 36}\oplus {\bf 42}$ of $USp(8)$, where
${\bf 36}$ is the adjoint of $USp(8)$. Therefore, the scalar fields are in the
${\bf 42}$, the fully anti-symmetric symplectic traceless four-tensor of $USp(8)$. The tangent vector of the coset $P_{\mu abcd}$ is defined as the totally antisymmetric component of the derivative $\tilde{V}^{AB} _{cd} D_\mu V_{AB} ^{ab}$ where tilde denotes the inverse of the vielbein $V$. 

The fields in the ungauged theory are governed by the Lagrangian:
\begin{equation}
    \begin{split}
        (16\pi G_5) e^{-1} \mathcal{L} &= R + \frac{i}{2}\bar{\psi}^a _\mu \gamma^{\mu \nu \rho} D_\nu \psi_{\rho a} -\frac{1}{8} G_{AB,CD} F^{AB}_{\mu \nu} F^{\mu \nu CD} - \frac{i}{12}\bar{\lambda}^{abc} \gamma^{\mu}D_{\mu } \lambda_{abc} - \frac{1}{6} P_{\mu abcd}P^{\mu abcd} \cr
        & -\frac{i}{3\sqrt{2}} P_{\mu abcd}\bar{\psi}_\mu ^a \gamma^{\rho} \gamma^\mu \lambda^{bcd} -\frac{i}{8}F_{\mu \nu}^{ab} \left(\bar{\psi}_a ^\rho \gamma_{[\rho} \gamma^{\mu \nu} \gamma_{\sigma]} \psi_b ^\sigma + \frac{1}{\sqrt{2}}\bar{\psi}_\rho ^c \gamma^{\mu \nu} \gamma^\rho \lambda_{abc} \right. \cr
        &\quad \left. + \frac{1}{2}\bar{\lambda}_{acd} \gamma^{\mu \nu} \lambda_b ^{\ cd}\right) + \frac{1}{24}e^{-1} \epsilon^{\mu \nu \rho \sigma \lambda} (F_{\mu \nu})^a _{\ b} (F_{\rho \sigma})^b _{\ c} (A_\lambda)^c _{\ a} ~. 
    \end{split}
\end{equation}
Here and in any subsequent Lagrangian, we do not include four-fermion terms. That is sufficient for our purposes, because we ultimately study quadratic fluctuations around a bosonic background. 
The determinant of the metric is $e = \sqrt{-\det g_{\mu \nu}}$. We mostly opt for $\text{USp}(8)$ indices, which are raised and lowered with the use of the symplectic matrix $\Omega_{ab}$, but we highlight the dependence of the gauge kinetic term on scalar fields by writing it in the $\text{E}_{6(6)}$ form 
\begin{equation}
\label{eqn:GABCDdef}
G_{AB,CD} = V_{ABab}\Omega^{ac}\Omega^{bd}V_{CDcd}~.
\end{equation}
$D_\mu$ is the $\text{USp}(8)$-covariant derivative that differs from the familiar covariant derivative in curved spacetime by a $\text{USp}(8)$ connection $Q_{\mu a} ^{\ b}$ that serves to project fields on to their physical components. We follow the conventions and notation established in \cite{Gunaydin:1985cu}. 

\subsection{From Ungauged to Gauged Supergravity}
\label{sec:ungaugetog}
Gauged supergravity is obtained from its ungauged progenitor
by gauging an $SU(4)$ subgroup of the global symmetry $\text{USp}(8)$ \cite{Gunaydin:1984qu,Gunaydin:1985cu}.
The spectrum of the resulting theory follows easily from the branching rule ${\bf 8}\to {\bf 4} + {\bf \bar 4}$ of the embedding $SU(4)\subset\text{USp}(8)$: 
\begin{itemize}
    \item The $\mathbf{8}$ gravitini into $\mathbf{4}+\mathbf{\bar{4}}$.
    \item The $\mathbf{27}$ vectors into $\mathbf{15}$ vectors and $\mathbf{6}+\mathbf{\bar{6}}$ antisymmetric tensors. 
    \item The $\mathbf{48}$ gaugini into $\mathbf{4}+\mathbf{\overline{4}}+\mathbf{20}+\mathbf{\overline{20}}$.
    \item The $\mathbf{42}$ scalars into $\mathbf{20'}+\mathbf{10}+\mathbf{\overline{10}}+\mathbf{1}+\mathbf{\bar{1}}$. 
\end{itemize}
However, the detailed structure of the gauged theory depends greatly on the nonlinear realization of a $SL(6,\mathbb{R})\times SL(2,\mathbb{R})$ subgroup of $\text{E}_{6(6)}$. The $SU(4)$ R-symmetry arises as the maximal compact subgroup $SU(4)\cong SO(6)\subset SL(6)$. 

This embedding is straightforward for the vector fields of the original ungauged theory. The fundamental $\text{E}_{6(6)}$ indices $AB$ decompose into $SL(6,\mathbb{R})$ indices $I,J,...$ and $SL(2,\mathbb{R})$ indices $\alpha, \beta,...$. For transformations under its compact subgroup, the fundamental of $SL(6,\mathbb{R})$ with index $I$ can be identified with the fundamental of $SO(6)$ with index $i$. As an example of this chain of identifications, the ${\bf 27}$ gauge field strengths $F^{ab}_{\mu\nu}$ in the ungauged theory reduce to $\mathbf{15} \ F^{ij} _\mu$ and $\mathbf{6}+\mathbf{\bar{6}} \ B^{i\alpha} _{\mu \nu}$ in the gauged theory. 

The analogous decompositions on the $27$-bein $V_{AB} ^{ab}$ make the $SO(6)$
content of the scalar fields more explicit\footnote{These formulae do not account for the full non-linear structure of the theory. The notation is simplified relative to \cite{Gunaydin:1985cu} but sufficiently precise for our purposes.}: 
\begin{align}
\label{eq: Vijab def}
    V_{ij} ^{\ ab} &= \frac{1}{8}(\Gamma_{kl})^{ab} S_i^k S_j^l + \frac{1}{4}(\Gamma^{k\alpha})^{ab}\varphi_{ijk\alpha} ~, \\ 
\label{eq: ViAab def}
    V_{i\alpha} ^{\ ab} &= \frac{1}{4\sqrt{2}} (\Gamma^{kl})^{ab} \varphi_{ikl\alpha} + \frac{1}{2\sqrt{2}} (\Gamma_{k\beta})^{ab} S^k _i \Lambda_\alpha ^{\beta}~. 
\end{align}
Here: 
\begin{itemize}
    \item The $\mathbf{20}'$ scalars $S^i _j$ parametrize the coset $SL(6)/SO(6)$ and can be identified with reparametrizations on the $S^5$. They are symmetric and traceless under $SO(6)$~. 
    We will also present these scalars as a symmetric matrix $T_{ij} = (SS^\top)_{ij}$ with unit determinant.
    \item The complex $\mathbf{1}_c$ scalars $\Lambda^\alpha _\beta$ parametrize the coset $SL(2)/U(1)$ and can be identified with the fundamental scalar field in ten-dimensional type IIB supergravity. They  are symmetric and traceless under $SL(2)$. 
    \item The $\mathbf{10}_c$ scalars $\varphi_{ijk\alpha}$ form a doublet under $SL(2)$. For $SO(6)$ they are real, totally antisymmetric in $i$, $j$ and $k$, and satisfy a self-duality relation, so they form a ${\bf 10}$. 
    \item $(\Gamma_i)^{ab}$ are the $8\times 8$ Euclidean gamma matrices of $SO(6)$. They are supplemented by a $(\Gamma_0)^{ab}$ that can be identified with the symplectic matrix $\Omega^{ab}$. The $SL(2)$ family $\Gamma_{i\alpha}$ are $\{ \Gamma_i, \Gamma_{i0} \}$. The $\Gamma_{ij} $ denotes the anti-commutator $\tfrac{1}{2}[\Gamma_i, \Gamma_j]$. 
\end{itemize}
The algebra of the $\Gamma$-matrices make the dependence of the gauge kinetic terms \eqref{eqn:GABCDdef} on the scalar fields introduced in \eqref{eq: Vijab def} more explicit: 
\begin{align}
G_{ij,kl} & = \frac{1}{4} (\delta_{mp}\delta_{nq}- \delta_{mq}\delta_{np})S^m_i S^n_j S^p_k S^q_l +\varphi_{ijm\alpha}\varphi_{klm\alpha} = \frac{1}{2}T^{-1}_{ik}T^{-1}_{jl}+\varphi_{ijm\alpha}\varphi_{klm\alpha}~.
 \end{align} 
 
The $27$-bein $V_{AB}^{~~ab}$ also define an auxiliary tensor:  
\begin{equation}
    \label{eq: Wabcd def}
    W_{abcd}= \epsilon^{\alpha \beta} \delta^{IJ}V_{I\alpha ab} V_{J\beta cd} ~,
\end{equation}
with the contractions:
\begin{align}
    \mathcal{T}_{ab} &= \frac{15}{4}W^c _{\ acb} \ , \ \mathcal{A}_{abcd}= -3W_{a[bcd]|} ~.
     \label{eq: TAabcd def}
  \end{align}  
The vertical bar in the subscript indicates subtraction of all symplectic traces. Our convention is that antisymmetrizations have coefficients $\frac{1}{n!}$. For example, $X_{[a b]} = \frac{1}{2}(X_{ab} - X_{ba})$.

To manage the presentation of the ${\cal N}=8$ gauged supergravity Lagrangian, we separate it as 
\begin{equation} \label{eq: L gauged}
    (16\pi G_5) e^{-1} \mathcal{L} =  \mathcal{L}_{g+v} +\mathcal{L}_s + \mathcal{L}_f~. 
\end{equation}

$\mathcal{L}_{g+v}$ combines the Einstein-Hilbert Lagrangian with the vector terms, including the Chern-Simons term: 
\begin{equation}
\begin{split}
\label{eq: L gauged g+v}
& \mathcal{L}_{g+v}= R -\frac{1}{8} G_{ij.kl}F_{\mu \nu }^{\ ij} F^{\mu \nu \ kl} -\frac{1}{96}e^{-1} \epsilon^{\mu \nu \rho \sigma \tau} \epsilon_{ijklmn} \left(F_{\mu \nu} ^{ij} F_{ \rho \sigma} ^{kl} A_{\tau} ^{mn} + g F_{\mu \nu} ^{ij} A_{\rho} ^{kl} A_{\sigma} ^{mp} A_{\tau} ^{pn} \right. \\ 
&\left. + \frac{2}{5}g^2 A_{\mu} ^{ij} A_{\nu} ^{kp} A_{\rho} ^{pl} A_{\sigma} ^{mr} A_{\tau} ^{rn} \right)+\frac{1}{8g}e^{-1} \epsilon_{\alpha \beta} \epsilon^{\mu \nu \rho \sigma \tau} B_{\mu \nu} ^{i\alpha} D_\rho B_{\sigma \tau} ^{j\beta} -\frac{1}{4} G_{i\alpha \ j\beta}B_{\mu \nu \ i\alpha} B^{\mu \nu \ j\beta}~.
    \end{split}
\end{equation}
The kinetic term in the ungauged theory divided into kinetic terms for the ${\bf 15}$ vectors $A_{\mu\ ij}$ and a mass term for the $\mathbf{6}+\mathbf{\bar{6}}$ tensors $B_{\mu\nu \ i\alpha}$. The latter acquire first order kinetic terms from the Chern-Simons interactions in the ungauged theory. It may be clearer to express
the Lagrangian in form notation, especially for the Chern-Simons terms: 
\begin{equation}
\label{eq: Lg+v}
    \begin{split}
        \mathcal{L}_{g+v} &= R\star 1 - \frac{1}{8}T_{ik}^{-1} T_{jl}^{-1} F^{ij} \wedge \star F^{kl} -\frac{1}{4}\varphi_{ijm\alpha} \varphi_{klm\alpha}F^{ij} \wedge \star  F^{kl} -\frac{1}{48}\epsilon_{i_1 ... i_6}\Big(F^{i_1 i_2} \wedge F^{i_3 i_4} \wedge A^{i_5 i_6}  \\
    &\quad 
-g F^{i_1 i_2} \wedge A^{i_3 i_4} \wedge A^{i_5 j} \wedge A^{j i_6} + \frac{2}{5}g^2 A^{i_1 i_2} \wedge A^{i_3 j} \wedge A^{j i_4} \wedge A^{i_5 k} \wedge A^{k i_6}\Big) ~\\
    & +\frac{1}{8g}e^{-1} \epsilon_{\alpha \beta} \epsilon^{\mu \nu \rho \sigma \tau} B_{\mu \nu} ^{i\alpha} D_\rho B_{\sigma \tau} ^{j\beta} - \frac{1}{4}
    G_{i\alpha \ j\beta} B_{\mu \nu} ^{i\alpha} B^{\mu \nu i \alpha} ~.
   \end{split}
\end{equation}
We also made the kinetic terms for the ${\bf 15}$ vectors more explicit. They are inherited rather straightforwardly from the $SL(6)/SO(6)$ and $SL(2)/U(1)$ coset structures \cite{cvetivc2000consistent}. The kinetic terms that have become mass-term for $B^{i\alpha}_{\mu\nu}$ are not made explicit because, in the approximation we need, they will collapse to the identity. 

$\mathcal{L}_s$ collects the kinetic and potential terms associated with the scalars: 
\begin{equation}
    \begin{split}
        \label{eq: L gauged scalars} 
        \mathcal{L}_s &= -\frac{1}{6}P_{\mu abcd} P^{\mu abcd} + \frac{1}{6}g^2 \left(\frac{64}{225}(\mathcal{T}_{ab})^2 - (\mathcal{A}_{abcd})^2\right) ~. 
    \end{split}
\end{equation}
The full non-linear form of the scalar kinetic terms is very complicated. At quadratic order, the Lagrangian can be made more explicit as
 \begin{equation}
    \begin{split}
    \mathcal{L}_s &= -\frac{1}{4}T_{ij}^{-1} T_{kl}^{-1} DT_{li} \wedge \star DT_{jk}-\frac{1}{2} D \Lambda_{\alpha} ^{\ \beta} \wedge \star D \Lambda_{\alpha} ^{\ \beta}  - \frac{1}{2} D \varphi_{ijk\alpha} \wedge \star  D \varphi_{ijk\alpha}  -V\star1 
    ~, 
    \end{split}
\end{equation}
where the potential term generated by the gauging procedure can be decomposed as:
\begin{equation}
\label{eqn:N=8pot}
    V = \frac{1}{2}g^2 (2T_{ij} T^{ij} - (T^i_i)^2) -\frac{1}{4}g^2 \varphi_{ijk\alpha}\varphi^{ijk\alpha} ~. 
\end{equation}
The $\mathbf{1}_c$ scalars $\Lambda^\alpha _\beta$ are entirely absent from the potential. They have become massless scalars propagating in the background. 

The final term in the Lagrangian is $\mathcal{L}_f$ that collects all terms with fermions: 
\begin{equation}
\begin{split}\label{eq: L gauged fermions}
\mathcal{L}_f &=\frac{i}{2}\bar{\psi}^a _\mu \gamma^{\mu \nu \rho} D_\nu \psi_{\rho a}  - \frac{i}{12}\bar{\lambda}^{abc} \gamma^{\mu}D_{\mu } \lambda_{abc}  -\frac{i}{3\sqrt{2}} P_{\mu abcd}\bar{\psi}_\mu ^a \gamma^{\rho} \gamma^\mu \lambda^{bcd}  \\
&\quad-\frac{i}{8}F_{\mu \nu}^{ab} \left(\bar{\psi}_a ^\rho \gamma_{[\rho} \gamma^{\mu \nu} \gamma_{\sigma]} \psi_b ^\sigma + \frac{1}{\sqrt{2}}\bar{\psi}_\rho ^c \gamma^{\mu \nu} \gamma^\rho \lambda_{abc}+ \frac{1}{2}\bar{\lambda}_{acd} \gamma^{\mu \nu} \lambda_b ^{\ cd}\right)  \\ 
&\quad +\frac{2i}{15}g \mathcal{T}_{ab} \bar{\psi}^a _\mu \gamma^{\mu \nu} \psi^b _\nu -\frac{i}{6\sqrt{2}} g \mathcal{A}_{dabc}\bar{\lambda}^{abc} \gamma^\mu \psi^d _\mu - ig \bar{\lambda}^{abc}\left(\frac{1}{2}\mathcal{A}_{bcde}-\frac{1}{45}\Omega_{bd} \mathcal{T}_{ce}\right)\lambda _a ^{\ de}~.
 \end{split}
\end{equation}
The dependence of the fermion terms on scalar fields depends greatly on the notation introduced in \eqref{eq: TAabcd def}.

After gauging, the covariant derivatives $D_\mu$ in various parts of the Lagrangian incorporate minimal coupling to the $SU(4)$ gauge field $A_\mu ^{ij}$. The covariant derivatives acting on the various bosonic fields are given by:
\begin{align}
\label{eq: 20 covD}
        DT_{ij} &= dT_{ij} + g(A^{ik} T_{kj} + A^{jk} T_{ki}) ~,\\ 
\label{eq: 1c covD}
    D \Lambda_\alpha ^{\ \beta} &= d \Lambda_\alpha ^{\ \beta} ~, \\ 
\label{eq: 10c covD}
    D \varphi_{ijk\alpha} &= d \varphi_{ijk\alpha} +gA_{l [i} \varphi_{jk]l\alpha} ~, \\ 
\label{eq: A covD}
        F^{ij} &= dA^{ij} + gA^{ik} \wedge A^{kj} ~, \\ 
\label{eq: B covD}
        D_\mu B^{i\alpha} _{\rho \sigma} &= \partial_\mu  B^{i\alpha} _{\rho \sigma} + gA_\mu ^{ij} B^{j \alpha} _{\rho \sigma} ~. 
\end{align}
In these equations we omit the dependence on the $\text{USp}(8)$ connections $Q_{\mu a}^b$ for brevity. They incorporate non-linear corrections to derivatives along the scalar manifold which, because the background we ultimately focus on has constant scalars, play no role. 

\subsection{The Equations of Motion in ${\mathcal N}=8$ Gauged Supergravity}
After the overview of the gauged Lagrangian and its field content, we can list the equations of motions for all fields. 

We start with the bosonic equations \cite{Gunaydin:1985cu, cvetivc2000consistent}. The exact equations of motion for the gauge fields $A_{ij}$ are
\begin{equation} \label{eq: A exact eoms form}
\begin{split}
D(T_{ik}^{-1} T_{jl} ^{-1} \star F^{kl} ) &= -2g T_{k[i} ^{-1} \star DT_{j]k} -\frac{1}{8}\epsilon_{ijk_1 ... k_4 } F^{k_1 k_2} \wedge F^{k_3 k_4} ~, 
\end{split}
\end{equation}
and the tensor equations are
\begin{equation} \label{eq: tensor eom}
    D_{[\mu } B_{\nu \rho]} ^{\ i\alpha} -\frac{1}{12}g \epsilon^{\alpha \beta} \epsilon_{\mu \nu \rho \sigma \tau}V_{i\beta} ^{ab} V_{ab} ^{j\gamma} B_{j\gamma} ^{\sigma \tau}  =0~. 
\end{equation}
The equations of motion for the scalar fields $T^{ij}$ are: 
\begin{equation} \label{eq: T exact eoms form}
\begin{split}
D(T_{ik} ^{-1} \star DT_{kj} ) &= -2g^2 (2T_{ik} T_{jk} - T_{ij} T_{kk})\star 1 + T_{ik}^{-1} T_{lm} ^{-1}  F^{lk} \wedge \star F^{mj}\\
&\quad -\frac{1}{6}\delta_{ij} \left[-2g^2 (T_{kl}T_{kl}-T_{kk}^2)\star 1 +T_{pk}^{-1} T_{lm} ^{-1} F^{lk} \wedge \star F^{mp}\right] ~.
\end{split}
\end{equation}
These equations are traceless because of the $\det T = 1$ constraint on the $T_{ij}$.
The equations of motion for the remaining scalar fields, $\Lambda_\alpha ^{\ \beta}$ and $\varphi_{ijk\alpha}$ are:
\begin{align}
\label{eq: 1c eom}
    \Box \Lambda_\alpha ^{\ \beta} &= 0 ~, \\ 
\label{eq: 10c eom}
    D_\mu D^\mu \varphi_{ijk\alpha} -3g^2 \varphi_{ijk\alpha} -\frac{1}{2}F^{lm} _{\mu \nu}  F^{\mu \nu [ij}\varphi_{k]lm\alpha} &= 0 ~.  
\end{align}
Finally, we have the fermionic equations: 
\begin{align} \label{eq: psi eom}
    &\gamma^{\mu \nu \rho} D_\nu \psi_{\rho a} -\frac{1}{3\sqrt{2}}  P_{\nu abcd} \gamma^\nu \gamma^\mu \lambda^{bcd} -\frac{1}{4}F_{ab} ^{\rho \sigma} \gamma^{[\mu} \gamma_{\rho \sigma} \gamma^{\nu]} \psi_\nu ^b -\frac{1}{8\sqrt{2}} F_{\rho \sigma} ^{bc} \gamma^{\rho \sigma} \gamma^\mu \lambda_{abc} \nonumber \\
    &\quad +\frac{4}{15}g \mathcal{T}_{ab} \gamma^{\mu \nu} \psi_\nu ^b -\frac{1}{6\sqrt{2}} g\mathcal{A}_{abcd}\gamma^\mu \lambda^{bcd} =0 ~, \\ 
    \label{eq: lambda eom}
    &\frac{1}{6}\gamma^\mu D_\mu \lambda_{abc} -\frac{1}{3\sqrt{2}}P_{\nu abcd}\gamma^\mu \gamma^\nu \psi^d _\mu -\frac{1}{8\sqrt{2}}\gamma^\rho \gamma_{\mu \nu} \psi_{\rho [a} F_{bc]|} ^{\mu \nu} \nonumber\\ 
    &\quad +\frac{1}{8}\gamma_{\mu \nu} \lambda^d _{\ [ab} F_{c]|d} ^{\ \mu \nu} +\frac{1}{6\sqrt{2}}g\mathcal{A}_{dabc}
\gamma^\mu \psi_\mu ^d +g \left(\frac{1}{2}\mathcal{A}_{[ab} ^{\ ef} +\frac{1}{45}\delta_{[a} ^{\ e} \mathcal{T}_b ^{\ f} \right)\lambda_{c]|ef} = 0~.
\end{align}
The fermionic equations of motion simplify considerably when applied to fermionic fluctuations around a bosonic background with a simple structure. 

\section{The Kerr-Newman-AdS background}
\label{sec:KNAbknd}

In this section we review the Kerr-Newman-AdS black holes that are familiar 
solutions to Einstein-Maxwell-AdS theory. We show how they lift to solutions of $\mathcal{N}=8$ gauged supergravity. The detailed verification of the 
$\mathcal{N}=8$ equations of motion forms the basis for the study of 
fluctuations in subsequent sections. 

\subsection{Kerr-Newman-AdS as a Solution to Einstein-Maxwell-AdS Theory}
The Kerr-Newman-AdS black holes are, by definition, solutions to Einstein-Maxwell-AdS (EMAdS) theory. In five dimensions, the EMAdS Lagrangian is:
\begin{equation} \label{eq: EMAdS L}
    \mathcal{L}_\text{EMAdS} = \frac{1}{16\pi G_5} \left((R+12g^2)\star 1 -\frac{3}{2}F \wedge \star F - F\wedge F \wedge A \right) ~. 
\end{equation}
The cosmological constant $-6g^2$ is related to the inverse of the AdS length scale as $\ell_5=g^{-1}$.
The gauge field strength is $F=dA$. The resulting equations of motion are the Maxwell equation and the trace-reversed Einstein equation: 
\begin{align}\label{eqn:EMAdSeom}
    d\star F + F \wedge F &= 0 ~, \\
    R_{\mu \nu} + 4g^2 g_{\mu \nu} &= \frac{3}{2}F_{\mu \lambda} F^{\lambda}_{\ \nu} -\frac{1}{4}g_{\mu \nu} F_{\lambda \rho} F^{\lambda \rho} ~. 
\end{align}
We will show that {\it any} solution to these equations can be recast as a solution to ${\cal N}=8$
gauged supergravity. 

In some situations, we will need explicit solutions, and then
we consider the Kerr-Newman-AdS black holes with mass $M$, electric charge $Q$, and a single angular momentum $J$. They are parametrized by three variables $(m,q,a)$ as: 
\begin{align}
\label{eqn:Mdef}
    M &= \frac{\pi}{4G_5} \frac{m(3 + a^2 g^2)+4qa^2 g^2}{\Xi^3} ~, \\ 
    \label{eqn:Qdef}
    Q &= \frac{\pi}{4G_5} \frac{q}{\Xi^2} ~, \\ 
    \label{eqn:Jdef}
    J &= \frac{\pi}{4G_5} \frac{2ma+qa(1+a^2 g^2)}{\Xi^3} ~,
\end{align}
where $\Xi = 1-a^2 g^2$. The more general family with two independent angular momenta \cite{chong2005general} is not considered in this article. 
The geometry and gauge field of the black holes are:
\begin{align}
\label{eq: KNAdS geom}
    ds^2 &= -\frac{ [(1+g^2 r^2)\rho^2 dt+2q\nu]dt}{\Xi \rho^2} + \frac{2q\nu^2}{\Xi \rho^2} + \frac{f}{\Xi^2 \rho^4}\left(dt-\nu\right)^2 \nonumber \\ 
    &\quad +\frac{\rho^2 dr^2}{\Delta_r} + \frac{\rho^2 d\theta^2}{\Xi} + \frac{r^2 + a^2}{\Xi} \left( \sin^2 \theta d\phi ^2 + \cos^2 \theta d\psi ^2\right) ~, \\
\label{eq: KNAdS A pot}
    A&= \frac{3q }{\Xi \rho^2} \left(dt-\nu\right) ~, 
\end{align}
where: 
\begin{equation}
\begin{split}
\label{eq: def geom variables}
    \Delta_r &= \frac{(r^2+a^2)^2(1+g^2 r^2)+q^2 +2a^2 q}{r^2}-2m  ~, \ \rho^2 = r^2 + a^2 ~,\\
    \nu&= a\left( \sin^2 \theta d\phi + \cos^2 \theta d\psi\right) ~,  \quad  f = 2m\rho^2 -q^2 +2a^2 g^2 q \rho^2 ~.
\end{split}
\end{equation}
The thermodynamic potentials of the black hole are given in the parametric form
\begin{align}
\label{eqn:Sdef}
S & =  \frac{\pi^2}{2G_5} \frac{(r^2_++a^2)^2 + a^2 q}{\Xi^2 r_+}~,\\
\label{eqn:Tdef}
T & =  \frac{r^4_+ (1 + 2g^2 (r_+ ^2 + a^2)) - (q+a^2)^2}{2\pi r_+ ((r^2_+ + a^2)^2 + a^2 q)}~,\\
\label{eqn:Omdef}
\Omega & =  a \frac{(r^2_++a^2) (1 + g^2 r^2_+) + q}{(r^2_++a^2)^2 + a^2 q}~, \\
\label{eqn:Phidef}
\Phi & =  \frac{qr^2_+}{(r^2_++a^2)^2 + a^2 q} ~,
\end{align}
Here the coordinate position $r_+$ of the black hole horizon is the largest solution to the $\Delta_r=0$ condition: 
\begin{equation}
\label{eqn:horeq}
(1+g^2 r^2_+)(r^2_++a^2)^2 + q^2 + 2a^2 q - 2mr^2_+=0~.
\end{equation}

\subsection{EMAdS as a Consistent Trunction of $\mathcal{N}=8$ Gauged Supergravity}

The field content of the Einstein-Maxwell-AdS theory \eqref{eq: EMAdS L} is just
gravity and a single gauge field $A_\mu$. Starting from any solution to this simple theory, we want to construct a solution to gauged ${\cal N}=8$ supergravity \eqref{eq: L gauged}. To do so, we must: 
\begin{enumerate}
    \item 
    Specify all fields in ${\cal N}=8$ SUGRA in terms of the fields in EMAdS. 
    \item 
    Show that the fields specified this way solve their equations of motion in ${\cal N}=8$ SUGRA, provided only that gravity and the single gauge field $A_\mu$ satisfy the equations of motion in EMAdS \eqref{eqn:EMAdSeom}.
\end{enumerate}

For some fields this is very simple to do. First of all, the fermions appear quadratically in the supergravity action \eqref{eq: EMAdS L}, so their equations of motion are linear in the fields. Therefore, we can simply specify that all fermion fields vanish, and this is sufficient to satisfy their equations of motion. 
Similarly, we will specify that the scalars $\Lambda_\alpha ^\beta$ in 
$\mathbf{1}_c$, the scalars $\varphi_{ijk\alpha}$ in $\mathbf{10}_c$, and the 
tensors in $\mathbf{6}+\mathbf{\bar{6}}$ all vanish. These fields transform non-trivially under $SL(2)$. This symmetry is respected by gauged supergravity, so all terms in the action are invariant under $SL(2)$, and this in turn means these fields cannot appear linearly in the action. This claim is of course easy to verify explicitly, by inspection of \eqref{eq: L gauged}. The result shows these fields are analogous to fermions, in that they automatically satisfy their equations motion if they vanish. Summarizing so far, we have specified
\begin{equation}
\label{eq: KNAdS background fields1}
    \Lambda_\alpha ^\beta = \varphi_{ijk\alpha} = B^{i\alpha} _{\mu \nu} = \psi^a _\mu = \lambda_{abc} = 0 ~, 
\end{equation}
and shown that these trivial values satisfy the equations of motion 
of ${\cal N}=8$ gauged supergravity.

It remains to specify the $\mathbf{20'}$ scalars $T_{ij}$ and the $\mathbf{15}$ vectors $A_{\mu ij}$ so their equations of motion (\ref{eq: A exact eoms form}-\ref{eq: T exact eoms form}) are satisfied, along with Einstein's equation. 
We want to allow non-trivial values of the gauge field $A_\mu$ that persists in the EMAdS theory, so generally the right hand sides of these equations do not simply vanish. That makes the task challenging. We specify:
\begin{equation}\label{eq: KNAdS background fields2}
    T_{ij} = \delta_{ij} \ , \ A_{\mu ij} = A_\mu \Omega_{ij} ~, 
\end{equation}
where $\delta_{ij}$ is the Kronecker delta, and $\Omega^{ij}$ is the symplectic matrix:
\begin{equation} \label{eq: Omega def}
    \Omega^{ij} = \begin{pmatrix}
        0 & 1 & & & & \\ 
        -1 & 0 & & & &  \\ 
        & & 0 & 1 & & \\ 
        & & -1 & 0 & & \\ 
        & & & & 0 & 1 \\ 
        & & & & -1 & 0 
    \end{pmatrix} ~. 
\end{equation} 
The scalar fields $T_{ij}=\delta_{ij}$ ensure the cancellation of the terms on the right hand sides of \eqref{eq: T exact eoms form} that only depend on scalar fields: the covariant derivative $DT_{ij}$ may be non-trivial, but the combination that appears does vanish. Also, the coefficients and the tensor structures of the $T^2$-terms are such that these terms cancel among themselves. 

The ${\cal N}=8$ gauge fields $A_{\mu ij}$ are specified in terms of the single Maxwell gauge field $A_\mu$ using the symplectic matrix \eqref{eq: Omega def}. This embedding is such that the $F^2$-terms in the equations of motion for $T_{ij}$ vanish. This is the consistency of specifying constant scalars. Moreover, it is such that the equation of motion for $A_{\mu ij}$ reduces to the index-less equation $d\star F = - F\wedge F$. 

The normalization $T_{ij}=\delta_{ij}$ is such that the ${\cal N}=8$ scalar potential 
$V$ \eqref{eqn:N=8pot} reduces to twice the $\text{AdS}_5$ cosmological constant: $V=-12g^2$ after truncation. Similarly, the normalization in the embedding $A_{\mu ij} = A_\mu \Omega_{ij}$ ensures that the ${\cal N}=8$ gauge kinetic and Chern-Simons terms in 
\eqref{eq: Lg+v} reduce to their analogues in EMAdS \eqref{eq: EMAdS L}, with the correct coefficient $\tfrac{3}{2}$.

In summary, our consistent truncation of $\mathcal{N}=8$ gauged supergravity to
the Einstein-Maxwell-AdS theory is given by (\ref{eq: KNAdS background fields1}-\ref{eq: KNAdS background fields2}). The manner in which the $\mathcal{N}=8$ equations of motion are satisfied is important, because it forms the basis of our later discussion of fluctuations around the background, so we summarize it here: 
\begin{itemize}
    \item 
    The fermion equations of motion \eqref{eq: psi eom} are satisfied by vanishing fermions $\psi^a _\mu = \lambda_{abc} = 0$, because they are linear in the fermions.  
    \item 
    The equations of motion for the tensors $B_{\mu \nu} ^{\ i\alpha}$ \eqref{eq: tensor eom} and the scalars $\Lambda_\alpha ^\beta$, $\varphi_{ijk\alpha}$
    (\ref{eq: 1c eom}-\ref{eq: 10c eom}) are similarly satisfied for vanishing fields, because they are linear.  
    \item The Maxwell equations for the $A^{ij} _\mu$  \eqref{eq: A exact eoms form} are satisfied nontrivially. 
        The source term $TDT$ vanishes for the constant background $T_{ij}=\delta_{ij}$. Further, the embedding $F_{ij} = F\Omega_{ij}$ simplifies the remaining $F_{ij}$ dependent terms so the equations reduce to the ordinary equation for a vector field with Chern-Simons couplings \eqref{eqn:EMAdSeom}.  
    \item The equation of motion \eqref{eq: T exact eoms form} for the scalars $T_{ij}$ is also satisfied nontrivially. The left hand side vanishes for constant background, but the source terms on the right-hand side require a precise cancellation in order to yield zero. 
    \item 
    With the matter fields we specify, the energy momentum tensor of ${\cal N}=8$ supergravity reduces to that of the EMAdS theory, so Einstein's equations reduce properly. 
\end{itemize}

\subsection{The Symmetry Breaking Pattern of the Consistent Truncation}
\label{sec:ssb}
The consistent truncation of ${\cal N}=8$ gauged supergravity identifies one of the {\bf 15} gauge fields $A_{\mu ij}$ as ``the" gauge field that persists in EMAdS. Conversely, that means the $14$ ``other" vector fields constitute extraneous matter. We can distinguish these fields algebraically, by utilizing the symplectic matrix $\Omega_{ij}$ \eqref{eq: Omega def} that characterizes the embedding \eqref{eq: KNAdS background fields2}. 
The {\bf 15} gauge fields $A_{\mu ij}$ are antisymmetric in the $SO(6)$ indices $i, j$. ``The" gauge field of EMAdS is identified by the symplectic trace $\Omega^{ij}A_{\mu ij}$. Conversely, the $14$ ``other" vector fields are in the $\Omega$-traceless part of $A_{\mu ij}$.

For a more general analysis, recall that fields in the theory are organized into representations of the $SU(4) \cong SO(6)$ $R$-symmetry. However, only the generators that leave $\Omega_{ij}$ invariant are symmetries also in the black hole background. The symmetries identified this way form the subgroup $SU(3)\times U(1)\subset SU(4)$. 
Fields transforming according to the vector ${\bf 8}$ of $\text{USp}(8)$ prior to gauging, branch into ${\bf 4}\oplus\bf{\bar{4}}$ under $SU(4)$ due to gauging, and then 
onto $({\bf 3}_1\oplus{\bf 1}_{-3})$ plus the complex conjugates 
$({\bf \bar 3}_{-1}\oplus{\bf 1}_{3})$ under $SU(3)\times U(1)$. The normalization of the $U(1)$ charge is set by a convention, but the contributions from the two terms in $SU(3)\times U(1)$ must cancel because $SU(4)_R$ has no overall $U(1)$. 

Following this branching rule, we can use the known $SU(4)_R$ field content, summarized in the beginning of subsection \ref{sec:ungaugetog},
to deduce the charges under the $SU(3)\times U(1)$ that is preserved by the black hole: 
\begin{itemize}
    \item The $\mathbf{8}$ gravitini have $\mathbf{4}+\mathbf{\bar{4}}$ $SU(4)_R$ charges which, in the black hole background, further branch into $\mathbf{3}_{1}\oplus\mathbf{1}_{-3} + \text{c.c.}$ under $SU(3)\times U(1)$. 
   \item The $27$ gauge fields arise from 
   $({\bf 8}\otimes {\bf 8})_{\rm as}=[ ({\bf 4}\otimes {\bf \bar 4}) \oplus ({\bf 4}\otimes {\bf \bar 4})]_{\rm as}$, minus the $USp(8)$ trace. Decomposing the 
   $({\bf 4}\oplus {\bf 4})_{\rm as} = {\bf 6}$ 
   under the $SU(3)\times U(1)$ subgroup, we conclude that the $\mathbf{6}_c$ tensors break into $\mathbf{3}_{-2} \oplus \mathbf{\bar{3}}_{2}+ \text{c.c.}$. 
  
   On the other hand, multiplying out ${\bf 4}\otimes {\bf \bar 4}=(\mathbf{3}_{1} \oplus\mathbf{1}_{-3}  )\otimes(\mathbf{\bar 3}_{-1} \oplus \mathbf{1}_{3} )$ minus a trace, we find that, in the black hole background, the $\mathbf{15}$ vectors branch into a $\mathbf{1}_0$, identified as ``the" vector in the EMAdS theory, and $\mathbf{8}_0 \oplus \mathbf{3}_{4} \oplus \mathbf{\bar{3}}_{-4}$. 
   
    \item The $SU(4)_R$ content of the $\mathbf{48}$ gaugini 
    is $\mathbf{4}$ $\lambda_{ABC}$ and
    $24-4=\mathbf{20}$ $\delta^{B{\bar C}}$-traceless $\lambda_{AB{\bar C}}$, as well as their complex conjugates. 
    The first contribution is equivalent to ${\bf \bar 4}$, which has $SU(3)\times U(1)$ content ${\bf \bar 3}_{-1}\oplus\mathbf{1}_{3}$. There is also a complex conjugate. 
    
    To work out the second contribution, recall again that $({\bf 4}\otimes {\bf 4})_{\rm as}={\bf 6}$ branches into $\mathbf{3}_{-2} \oplus \mathbf{\bar{3}}_{2}$. Further tensoring with ${\bf \bar 4}={\bf\bar 3}_{-1}\oplus\mathbf{1}_{3} $ then gives the $SU(3)\times U(1)$ content
        $\mathbf{8}_{-3}\oplus \mathbf{\bar 6}_{1}\oplus\mathbf{\bar 3 }_{5}\oplus\mathbf{3}_1$ and
       a ${\bf 4}={\bf 3}_{1}\oplus\mathbf{1}_{-3}$ that is removed. Again, there is also a complex conjugate.
    \item 
    The $\mathbf{10}_c$ scalars $\varphi_{ijk\alpha}$ correspond to $SU(4)_R$ symmetric tensors $\varphi_{AB} + {\rm c.c.}$ that branch to $\mathbf{6}_{2} \oplus \mathbf{3}_{-2} \oplus \mathbf{1}_{-6}+ {\rm c.c.}$. 

  The $\mathbf{20'}$ scalars are in the symmetric traceless $SO(6)$ representation. Recalling again the
  $SU(4)\to SU(3)\times U(1)$ branching ${\bf 6} \to \mathbf{3}_{-2} \oplus \mathbf{\bar{3}}_{2}$, 
  we find
\begin{equation}
    \left(({\mathbf{3}}_{-2} \oplus \mathbf{\bar 3}_2) \otimes ({\mathbf{3}}_{-2} \oplus \mathbf{\bar 3}_2)\right)_\text{sym, tr} = (\mathbf{3}_{-2} \otimes \mathbf{3}_{-2})_\text{sym}
    \oplus (\bar{\mathbf{3}}_2 \otimes \bar{\mathbf{3}}_2)_\text{sym} \oplus (\bar{\mathbf{3}}\otimes \mathbf{3})_\text{tr}  =\mathbf{6}_{-4} \oplus \mathbf{\bar{6}}_4 \oplus \mathbf{8}_0~. 
\end{equation}
\end{itemize}
The group assignments determined by the symmetry breaking pattern in this way require that fields with distinct $SU(3)$ quantum numbers cannot mix at all, and fields with distinct $U(1)$ assignments can only mix through couplings with a compensating background field strength. These principles will be realized by the explicit equations of motion for fluctuations derived in the next section. The group theory assignments found here are summarized in Table \ref{tab:scallabel} and \ref{tab:veclabel}, along with other aspects of the equations of motion. 


The organization of fluctuations around a black hole by the quantum numbers of broken global symmetries is highly constraining. In the AdS vacuum, algebraic methods are sufficient to fully diagonalize the equations of motion and compute the conformal dimensions \cite{Gunaydin:1984fk}. In some favorable circumstances, similar results apply in the AdS$_2$ near horizon geometry of an extremal black hole \cite{Maldacena:1998bw,Larsen:1998xm,deBoer:1998kjm}. With these considerations in mind, we exploit the global symmetries and assemble the fluctuating fields into the unique blocks that are consistent with the global symmetries: 

\begin{itemize}
\item
The {\bf gravity} block, $8+8$ d.o.f.: gravity, 2 gravitini, graviphoton. They are singlets under $SU(3)$. The bosons are neutral under $U(1)$, but the gravitini have charges $\pm 3$. In AdS$_5$, the graviton has $\Delta=4$, the gravitini $\Delta=\frac{7}{2}$, and the vector $\Delta = 3$.

\item
The {\bf massive gravitino} block, $6 \times (6+6)$ d.o.f.: gravitino, $2$ tensors, $1$ chiralino. The gravitini are in ${\bf 3}_{1} + {\bf{\bar 3}_{-1}}$, the tensors in $2({\bf 3}_{-2} + {\bf{\bar 3}_{2}})$, and the chiralini in ${\bf 3}_{-5} + {\bf{\bar 3}_{5}}$. In AdS$_5$, the gravitino has $\Delta=\frac{7}{2}$, the tensors $\Delta=3$, and the chiralino $\Delta = \frac{5}{2}$.

\item
The {\bf gauge field} blocks: $8\times (4 + 4)$ d.o.f.: $1$ gauge field, $2$ chiralini, $1$ scalar. All are in the adjoint ${\bf 8}$ of $SU(3)$ and have $\Delta=3, \frac{5}{2}, 2$ in AdS$_5$.
The bosons are neutral under $U(1)$, but the chiralini have charges $\pm 3$.

\item
The {\bf massive gauge field} blocks: $6 \times (4 + 4)$ d.o.f., $1$ gauge field, $2$ chiralini, and $1$ scalar. The gauge fields are in the fundamental (anti-fundamental) 
${\bf 3_4} \oplus {\bf{\bar 3}_{-4}}$, the chiralini are in ${\bf 3_1}\oplus {\bf {\bar 3}_{-1}}$, and the scalars are in ${\bf 3_{-2}} \oplus {\bf{\bar 3}_{2}}$.  

In AdS$_5$ these fields have dimensions $\Delta = 3, \frac{7}{2}, 3$.
The vectors and the scalars have the same dimension and, in the black hole background, it is artificial to distinguish them: they combine into a single massive multiplet. 

\item
The {\bf SU(3) tensor} blocks: $12 \times (2 + 2)$ d.o.f., each with $2$ scalars and $1$ chiralino. The scalars from the ${\bf 20}'$ are in ${\bf 6_{-4}} \oplus {\bf {\bar 6}_4}$, chiralini in ${\bf 6_{-1}}\oplus{\bf {\bar 6_{1}}}$, and the scalars from ${\bf 10}_c$ are in ${\bf 6_{2}} \oplus {\bf {\bar 6}_{-2}}$.
In AdS$_5$ these fields have dimensions $\Delta=2, \frac{5}{2}, 3$.

\item
The {\bf singlet} block: $(4 + 4)$ d.o.f., $4$ scalars and $2$ chiralini. They are all singlets ${\bf 1}$ under $SU(3)$ and have AdS$_5$ $\Delta = 3, \frac{7}{2}, 4$. The two
light scalars with  $\Delta=3$, from the ${\bf 10_c}$, are in ${\bf 1_6}$, ${\bf 1_{-6}}$. The chiralini are in ${\bf 1_3}$ and ${\bf 1_{-3}}$. The two heavy scalars with  $\Delta=4$ are in ${\bf 1_0}$, they were singlets all along. 
\end{itemize}
There is a total of $128+128$ d.o.f., as there should be. The claim of the block structure is that, at quadratic level around the Kerr-Newman background, the equations of motion cannot mix different blocks. The structure applies in the entire spacetime, also away from the horizon region. The pairing between fermions and bosons is inherited from the supersymmetry of the theory. 
It is consistent with supersymmetry that has global charge assignments ${\bf 1}_{\pm 3}$, but we do not presently assume that the black hole background preserves supersymmetry. It is interesting that, even at this general level, the entire symmetry structure fits nicely with expectations from spontaneous breaking of the $PSU(2, 2|4)$ superconformal symmetry to the $PSU(1, 1|3)$ that is respected by the subgroup that preserves $1/16$ supersymmetry \cite{Kinney:2005ej}.

\subsection{Kerr-Newman AdS Black Hole in ${\mathcal N}=2$ Supergravity}

We have discussed in detail how the Kerr-Newman-AdS black hole solutions to the EMAdS theory are also solutions to ${\cal N}=8$ supergravity, and we found the equations of motion for fluctuations around the black hole of the matter specified by the ${\cal N}=8$ theory. Many studies of supersymmetric black holes focus instead on ${\cal N}=2$ supergravity, so it is worthwhile to recast our results in that language. 

Minimal ${\cal N}=2$ supergravity is the extension of EMAdS that includes $2$ gravitini. This is minor, in our context, because the bosonic field content of these theories is identical. It is natural to study more general theories that couple ${\cal N}=2$ supergravity to $n_V$ ${\cal N}=2$ vector-multiplets so that, taking the gauge field in the supergravity multiplet into account, they have a total of $n_V+1$ vector fields and $n_V$ scalar fields. Even more general models further allow for $n_H$ ${\cal N}=2$ hyper-multiplets. To make contact with the results for ${\cal N}=8$ supergravity, we also allow for ${\cal N}-2$ massive gravitini multiplets. Our reasoning generalizes previous work on black holes in ungauged supergravity to AdS black holes \cite{Keeler:2014bra,Castro:2018hsc,Larsen:2018cts}. Similar questions were pursued in AdS$_4$ supergravity \cite{Castro:2021wzn}.

A particularly important special case of this general setup is the STU model that has $n_V=2$ and couplings between gravity and ${\cal N}=2$ vectors specified by a canonical prepotential that is often written using the letters $S$, $T$, and $U$. The Kerr-Newman-AdS black hole is known to be a solution to the STU-model \cite{Cveticˇ_Duff_Hoxha_Liu_Lü_Lu_Martinez-Acosta_Pope_Sati_Tran_1999, cvetivc2004charged, gutowski2004general}. We expect it is similarly a solution to more general ${\cal N}=2$ supergravity theories with $n_V$ vectors, $n_H$ hypers, and ${\cal N}-2$ gravitini, when prepotentials and moment-maps are picked appropriately. 
With this assumption, we can recast our results for perturbations in ${\cal N}=8$ gauged supergravity in terms of very general ${\cal N}=2$ theories:

\begin{itemize}
\item 
The {\bf gravity} block is associated with {\it minimal} ${\cal N}=2$ supergravity. 
\item 
    The ${\cal N}-2$ {\bf massive gravitino} blocks appear when supersymmetry is enhanced above ${\cal N}=2$. In ${\cal N}=8$ supergravity, they are duplicated $6$-fold. For fluctuations around Kerr-Newman AdS black holes in ${\cal N}=4$ supergravity, 
    these fluctuations should have multiplicity $2$. We refer to these multiplets as ``massive" because they correspond to supersymmetries that are broken. 
\item 
The {\bf gauge field} block appears when the ${\cal N}=2$ theory has gauge symmetry beyond ``the" $U(1)$ R-symmetry that is part of minimal ${\cal N}=2$ supergravity. Generally, there are $n_V$ ${\cal N}=2$ vector multiplets. We can interpret ${\cal N}=8$ supergravity as minimal ${\cal N}=2$ supergravity with $n_V=8$ vector multiplets that realize $SU(3)$ gauge symmetry. 

The scalars in the ${\cal N}=2$ gauge multiplet have $\Delta=2$ in the AdS$_5$ vacuum, and so they are on the boundary of the BF-bound. However, because they are neutral under the $U(1)$ charge that the black hole carries, they do not condense in the near horizon environment. This type of scalars were denoted $X$ in early, influential studies \cite{liu2007new, basu2010small, bhattacharyya2010small, dias2012hairy}.

\item 
The {\bf massive gauge field} block is associated with {\it broken} gauge symmetry. In ${\cal N}=8$ supergravity, the Kerr-Newman AdS black hole singles out a $U(1)\subset SU(4)_R$, and so it breaks the gauge symmetry $SU(4)_R\to SU(3)\times U(1)$. As noted already, in this environment, the fields that would be vectors and scalars in AdS$_5$ combine into a single massive vector field.  

\item 
The {\bf SU(3) tensors} are {\it hyper-multiplets} from the ${\cal N}=2$ point of view. The scalars in these multiplets have $\Delta=2$ and they are charged under the same $U(1)$ as the black hole. Therefore, our embedding of the Kerr-Newman AdS black hole into ${\cal N}=8$ supergravity is a holographic realization of superfluidity \cite{hartnoll2008building, hartnoll2008holographic, franco2010general, horowitz2011introduction} in a theory that is UV complete. That is significant, because it is regularly stated, even recently \cite{Sachdev:2023fim}, that such embeddings were not yet constructed convincingly. 

The $SU(3)$ gauge symmetry, and these fields being tensors under that symmetry, are particular to the ${\cal N}=8$ supergravity progenitor. On the other hand, we expect that the charge assignment under the $U(1)_R$ symmetry is fixed by ${\cal N}=2$ supersymmetry. 
Therefore, the realization we discuss is likely to be rather generic. 

The early studies referred to previously \cite{liu2007new, basu2010small, bhattacharyya2010small, dias2012hairy} denoted this type of scalars by $\varphi$.
 
\item 
The {\bf singlet} block is the {\it universal} hyper multiplet. It is associated with the {\it coupling constant} of the theory. 
\end{itemize}

In summary of this subsection: because of symmetries, we expect that the equations of motions for fluctuations that we study in ${\cal N}=8$ gauged supergravity, apply also to black holes in large classes of ${\cal N}=2$ supergravity theories. The identifications we identify will apply only at highly symmetric points in moduli space, but may serve as a valuable benchmark more generally.

\section{Fluctuations Around the Background}
\label{sec:fluc}

In the previous section, the Einstein-Maxwell-AdS theory was embedded into $\mathcal{N}=8$ gauged supergravity fields so that any solution to EMAdS, like
the Kerr-Newman-AdS black hole, can be interpreted as a solution to the $\mathcal{N}=8$ theory. In this section, we derive the explicit equations of motion for the quadratic fluctuations of $\mathcal{N}=8$ gauged supergravity around the background solution (\ref{eq: KNAdS background fields1}-\ref{eq: KNAdS background fields2}). 

We focus on the bosonic fluctuations and do not consider fluctuations within EMAdS itself, which involve gravity and one gauge field. Those were discussed in \cite{Castro:2018ffi, kolekar2018ad}. The remaining fields are $42$ scalars and $26$ vectors. At the quadratic level, the fluctuations of the $\mathbf{20'}$ scalars and $\mathbf{14}$ vectors experience rather elaborate couplings. The pseudoscalars $\mathbf{10}_c$ are simpler. The final fields are the $\mathbf{1}_c$ scalars and the $\mathbf{6}_c$ vectors, which are nearly trivial. 

\subsection{Decoupling $\mathbf{20}'$ scalars and $\mathbf{14}$ Vectors}
The $\mathbf{20}'$ scalars and $\mathbf{15}$ gauge fields satisfy equations of motion (\ref{eq: A exact eoms form}-\ref{eq: T exact eoms form}) that are coupled to one another. Moreover, these fields are non-vanishing in the background solution \eqref{eq: KNAdS background fields2}. We parametrize the fluctuations around the background as:
\begin{equation} \label{eq: T A flucts def}
    T_{ij} = \delta_{ij} + t_{ij} \ , \ A^{ij} = A\Omega^{ij} + a^{ij} ~.  
\end{equation}
The $T_{ij}$ form a unimodular matrix with respect to the identify matrix,
so the fluctuations $t_{ij}$ are $\delta$-traceless. The fluctuating gauge fields $a^{ij}$ are $\Omega$-traceless, so there are $14$ independent components of $a^{ij}$.

The equations of motion depend on the covariant derivatives, defined in 
\eqref{eq: 20 covD} and \eqref{eq: A covD}, that take into account both the curved background and the non-abelian gauge group $SU(4)_R$. The decomposition \eqref{eq: T A flucts def} gives:
\begin{equation} \label{eq: DT DA flucts}
\begin{split}
        DT_{ij} &= \underbrace{dt_{ij} + gA(\Omega^{ik}t_{kj} + \Omega^{jk} t_{ki})}_{Dt_{ij}} + \underline{g(a^{ik} t_{kj} + a^{jk} t_{ki})} ~, \\ 
        F^{ij} &= F\Omega^{ij} + \underbrace{da^{ij}+gA\wedge (\Omega^{ik} a^{kj} - a^{ik} \Omega^{kj})}_{f^{ij}=Da^{ij}} +\underline{ g a^{ik} \wedge a^{kj}}~. 
\end{split}
\end{equation}
The underlined terms are higher-order, they are neglible in the linearized equations of motion and quadratic terms in the Lagrangian. The covariant derivatives on the fluctuating fields are denoted by $Dt_{ij}$, $Da_{ij}$ and defined by these equations. 

After expansion to linear order, the equations of motion 
(\ref{eq: A exact eoms form}-\ref{eq: T exact eoms form}) become:
\begin{equation} \label{eq: eoms t f}
\begin{split}
    D \star D t_{ij} - 4g^2 t_{ij}  \star 1 +2\Omega^{il} t_{lm} \Omega^{mj} (F\wedge \star F) &= -F \wedge \star (f^{kj} \Omega^{ki} + f^{ki} \Omega^{kj}) ~, \\ 
    D \star  f^{ij} + F \wedge \star (\Omega^{ik} f^{kl}  \Omega^{lj}) &= D \left( (t_{ik}\Omega^{kj} + t_{jk}\Omega^{ki})\star F\right) ~. 
    \end{split}
\end{equation}
Comments:
\begin{itemize}
    \item The mass and source terms in the $t_{ij}$ equation are automatically $\delta$-traceless. This follows from properties of $t_{ij}$ and $f_{ij}$.
    \item The covariant derivatives on $t_{ij}$ and $a_{ij}$, defined in \eqref{eq: DT DA flucts}, encode minimal couplings to the background gauge field $A$. \item 
    Both equations also have non-minimal couplings that depend on the background field strength $F$, rather then the gauge field $A$. 
    \item The fields $t_{ij}$ and $a^{ij}$ do not have independent equations of motion, they mix.  
\end{itemize}

\subsection*{The $\Omega$-operation: Block Diagonalizing the Equations of Motion}
The equations of motion \eqref{eq: eoms t f} for the fluctuations $t_{ij}$ and $a_{ij}$ are tensors of $SO(6)$. The equations are in the symmetric traceless and antisymmetric representations, respectively. We can simplify these equations by exploiting the subgroup of $SO(6)$ that is respected by the black hole background. The symplectic matrix $\Omega$ introduced in \eqref{eq: Omega def} decomposes the ${\bf 20}'$ scalar fluctuations $t_{ij}$ as
\begin{equation}
    t_{ij} = t_{ij} ^+ + t_{ij} ^- ~,  
\end{equation}
where $t_{ij} ^\pm$ satisfy
\begin{equation}
\label{eq: Omega action tpm}
    \Omega^{im}t^\pm_{mn}\Omega^{nj} =\pm t^\pm _{ij}  ~.
\end{equation}
For the upper sign, the equation imposes $8$ conditions, and for the lower it imposes $12$. The decomposition of a generic symmetric traceless $t_{ij}$ becomes:
\begin{align} \label{eq: tm decomp}
    t_{ij} ^- &=  \frac{1}{2}\begin{pmatrix} t_{11} +t_{22} & t_{13}+t_{24} & t_{15}+t_{26} \\
     & t_{33}+t_{44} & t_{35}+t_{46} \\
     &  & t_{55}+t_{66}
    \end{pmatrix} \otimes I + \frac{1}{2}\begin{pmatrix}
        0 & t_{14}-t_{23} & t_{16}-t_{25} \\
         & 0 & t_{36}-t_{45} \\ 
         & & 0 
    \end{pmatrix} \otimes (i\sigma_2) \\ 
    \label{eq: tp decomp}
    t_{ij} ^+ &=  \frac{1}{2}\begin{pmatrix} t_{11} -t_{22} & t_{13}-t_{24} & t_{15}-t_{26} \\
     & t_{33}-t_{44} & t_{35}-t_{46} \\
     &  & t_{55}-t_{66}
    \end{pmatrix} \otimes \sigma_3+ \frac{1}{2}\begin{pmatrix}
        2t_{12} & t_{14}+t_{23} & t_{16}+t_{25} \\
         & 2t_{34} & t_{36}+t_{45} \\ 
         & & 2t_{56} 
    \end{pmatrix} \otimes \sigma_1 ~.
\end{align}
For typographical clarity the bottom triangular halves were omitted, since they anyway follow from $t_{ij}^\pm =t_{ji} ^\pm$. The $\sigma_i$ refer to the Pauli matrices. With this decomposition, the $20=21-1$ components of
the symmetric, $\delta$-traceless fluctuations $t_{ij}$ split into an $\Omega$-odd subset $t^-_{ij}$  with $8=9-1$ $\delta$-traceless components, and an $\Omega$-even subset $t^+_{ij}$ that has $12$ components. This makes explicit 
a decomposition $\mathbf{20} = \mathbf{8}+\mathbf{12}$ for the $t_{ij}$.

We similarly decompose the {\bf 14} $a^{ij}$ as
\begin{equation}
    a_{ij} = a_{ij} ^+ + a_{ij} ^- ~,  
\end{equation}
where $a_{ij} ^\pm$ satisfy: 
\begin{equation}
\label{eq: Omega action apm}
    \Omega^{im}a^\pm_{mn}\Omega^{nj} =\pm a^\pm _{ij}  ~.
\end{equation}
The decomposition of a generic antisymmetric, $\Omega$-traceless $a_{ij}$ becomes:
\begin{align}
\label{eq: am matrix decomp}
    a^{ij} _- &=  \frac{1}{2}\begin{pmatrix}
    0 & a^{13}+a^{24} & a^{15}+a^{26} \\
    & 0 & a^{35}+a^{46} \\ 
    & & 0
    \end{pmatrix} \otimes I + \frac{1}{2}\begin{pmatrix}
    2a^{12} & a^{14}-a^{23} & a^{16}-a^{25}\\
    & 2a^{34} & a^{36}-a^{45} \\ 
    & & 2a^{56} 
    \end{pmatrix} \otimes (i\sigma_2) \\ 
    \label{eq: ap matrix decomp}
    a^{ij} _+ &=  \frac{1}{2}\begin{pmatrix}
    0 & a^{13}-a^{24} & a^{15}-a^{26} \\
    & 0 & a^{35}-a^{46} \\ 
    & & 0
    \end{pmatrix} \otimes \sigma_3+ \frac{1}{2}\begin{pmatrix}
        0 & a^{14}+a^{23} & a^{16}+a^{25} \\ 
        & 0 & a^{36}+a^{45} \\ 
        & & 0
    \end{pmatrix} \otimes \sigma_1 ~,
\end{align}
where again the bottom triangular halves have been omitted. They now follow from $a_{ij}^\pm =-a_{ji} ^\pm$. 
For the gauge field fluctuations, the $14=15-1$ $\Omega$-traceless components of the antisymmetric $a^{ij}$ split into an $\Omega$-odd subset $a^-_{ij}$ with $8=9-1$ entries, and an $\Omega$-even subset 
$a^+_{ij}$ with $6$ entries. 

According to our decomposition, the $\Omega$-symmetric part
of the equations of motion \eqref{eq: eoms t f} become:
\begin{equation}
\label{eqn:tp eom}
    D \star D t_{ij} ^+ - 4g^2 t_{ij} ^+  \star 1 -2t_{ij} ^+ (F\wedge \star F) = 0 ~, 
\end{equation}
and
\begin{equation}   
\label{eqn:ap eom}
    D \star  f^{ij} _+ + F \wedge  f^{ij} _+ = 0~,\\ 
\end{equation}
where, based on the definitions \eqref{eq: Omega action tpm} and \eqref{eq: Omega action apm}, $D$ acts on $a_+$ and $t^+$ as:
\begin{align}
        Dt^+ _{ij} &= dt^+ _{ij} + 2gA \Omega^{ik} t_{kj} ^+ ~, \\  
        f_+ ^{ij} &= Da_+ ^{ij} = da_+ ^{ij} +2g\Omega^{ik} A \wedge a^{kj} _+ ~.  
\end{align}
There are $12$ independent scalar fields $t_{ij}^+$. They have the mass $m^2=-4g^2$ that is familiar from the mass of all ${\bf 20}'$ scalars in the $AdS_5$-background. This mass is at the boundary of the BF-bound in AdS$_5$. In the Kerr-Newman-AdS environment each of these $12$ scalar fields experiences a non-minimal $F_{\mu\nu}F^{\mu\nu}$ coupling to the background gauge field. 
The $6$ independent gauge fields $a^+_{ij}$ are massless, as expected for all ${\bf 15}$ gauge fields in AdS$_5$, but these $6$ gauge fields 
also experience a non-minimal coupling to the background field strength. 

The $\Omega$-antisymmetric part of the equations of motion \eqref{eq: eoms t f} are:
\begin{align}
\label{eqn:tm eom original}
        D \star D t_{ij} ^- - 4g^2 t_{ij} ^-  \star 1 +2t_{ij} ^- (F\wedge \star F) &=  2F \wedge \star (f^{ki} _- \Omega^{kj})~, \\
        \label{eqn:am eom original}
    D \star  (f^{ki}_-\Omega^{kj}) - F \wedge  (f^{ki} _-\Omega^{kj}) &= - 2D \left( t_{ij} ^-\star F\right) ~. 
    \end{align}
We note in particular that due to the definitions of $t^-$ and $a_-$ in\eqref{eq: Omega action tpm} and \eqref{eq: Omega action apm}, $D$ acts on $t^-$ and $a_-$ as:
\begin{equation}
    Dt^- _{ij} = dt^- _{ij} \ , \ f_- ^{ij} = Da_- ^{ij} = da_- ^{ij} ~. 
\end{equation}
The simplification identified at this point is that there are $8$ independent blocks of equations that all take identical form. Each block comprises a scalar $t^-$ and a gauge field $a^-_\mu$ that couple to one another in an unfamiliar way. 

\subsection{The Pseudo-Scalars}

With some abuse of language, we refer to the scalar fields in $\mathbf{10}_c$ as pseudo-scalars, because they are are odd under the $SO(6)$ rotation group of $S^5$. These fields satisfy \eqref{eq: 10c eom}, which we make more explicit as: 
\begin{align}
\label{eqn:DDvarphi}
   D^\mu D_\mu \varphi_{ijk\alpha} -3g^2 \varphi_{ijk\alpha} -\frac{1}{4}F^{\mu \nu} F_{\mu \nu} \Omega_{[ij}\varphi_{k]lm\alpha}\Omega^{lm} &= 0 ~.  
\end{align}
To digest these equations, two aspects must be analyzed: the minimal couplings contained in the gauge covariant derivative, and the coupling to $F_{\mu\nu}F^{\mu\nu}$ due to the dependence of the gauge kinetic terms on scalar fields. 

As in previous cases, the symmetrization prescription $\Omega_{[ij}\varphi_{k]lm\alpha}\Omega^{lm}$ permits a decomposition under the $\Omega$-operation. The symplectic pairing, of the form $(12), (34), (56)$, identifies $12$ field components $\varphi_{ijk\alpha}$ where the indices $ijk$ include one pair, and $8$ components where $ijk$ belong to distinct pairs. Among the latter, the symplectic form further assigns one complementary set, such as $135$ and $246$, as a subspace with positive sign. Importantly, only the former appears in the combination $\Omega_{[ij}\varphi_{k]lm\alpha}\Omega^{lm}$. Because of the overall self-duality condition on $\varphi_{ijk\alpha}$, these index structures identify groups of $6, 3, 1$ field components that satisfy identical field equations. 

In a more covariant approach, the quadratic term in the Lagrangian \eqref{eq: Lg+v} has index structure $\varphi_{ijm\alpha} \varphi_{klm\alpha} \Omega^{ij} \Omega^{kl}$. The symplectic form $\Omega^{ij}$ specified by the background gauge field selects components of $\varphi_{ijk\alpha}$ where two of the three indices are within one ``pair", corresponding to ${\bf 6}$ of $SU(3)$. This projection on the fields having no Pauli terms can be expressed covariantly as: 
\begin{align}
\label{eq: varphi1 zero pauli}
    \Omega_{[ij}\varphi_{k]lm\alpha} ^{(\mathbf{1})}\Omega^{lm} &= 0~, \\
\label{eq: varphi3 zero pauli}
\Omega_{[ij}\varphi_{k]lm\alpha} ^{(\mathbf{3})}\Omega^{lm} &= 0 ~.
\end{align}
The remaining $\varphi ^{(\mathbf{6})}$ with nonvanishing Pauli-like terms can be diagonalized and reorganized, amounting to a Lagrangian term:
\begin{equation}
\label{eq: varphi6 pauli L term}
    +\frac{1}{3}F_{\mu \nu} F^{\mu \nu} (\varphi_{ijm\alpha} ^{(\mathbf{6})})^2 ~,
\end{equation}
which, incorporating the $3!=6$ degeneracy in the contraction, amounts to a $-2F_{\mu \nu}F^{\mu \nu}$ Pauli-like mass term contribution to the scalar Lagrangian. 

It follows from these considerations that, when forming linear combinations of \eqref{eqn:DDvarphi} that diagonalize the $\Omega$-operation, only $1$ instance gives gauge coupling ``$+$" for all three indices, and one gives all ``$-$"; the remainder mixes the signs. For this reason, it is only the one ``special" field component that acquires charge assignments $\pm 3$, while $9$ components have eigenvalue $\pm 1$:
\begin{align}
    D_\mu \varphi_{ijk\alpha} ^{(\mathbf{1})} &= \partial_\mu \varphi_{ijk\alpha} ^{(\mathbf{1})} \pm 3g A_\mu \varphi_{ijk\alpha} ^{(\mathbf{1})}~,\\
    D_\mu \varphi_{ijk\alpha} ^{(\mathbf{3})} &= \partial_\mu \varphi_{ijk\alpha} ^{(\mathbf{3})} \pm g A_\mu \varphi_{ijk\alpha} ^{(\mathbf{3})}~,\\
    D_\mu \varphi_{ijk\alpha} ^{(\mathbf{6})} &= \partial_\mu \varphi_{ijk\alpha} ^{(\mathbf{6})} \pm g A_\mu \varphi_{ijk\alpha} ^{(\mathbf{6})}~.
\end{align}
In our index manipulations here, we were not careful about the signs imposed by the self-duality condition on $\varphi_{ijk\alpha}$. That is not needed for our explicit computation to confirm
the $SU(3)\times U(1)$ charge assignments $\mathbf{1}_{-6} + \text{c.c.}$ and $\mathbf{6}_2 + \mathbf{3}_{-2} + \text{c.c.}$ derived in section \ref{sec:ssb} using symmetry principles. 
\subsection{The Final Bosons}

The remaining bosonic fields are much simpler, since they barely depend on the background at all. The  singlet scalars $\Lambda_\alpha ^{\ \beta}$ are massless and neutral, according to \eqref{eq: 1c eom}:
\begin{align}
\label{eq: Lambda eom}
    \Box \Lambda_\alpha ^{\ \beta} &= 0 ~.  
\end{align}
The designation of the ``complex" representation $\mathbf{1}_c$ refers to this pair of fields transforming under the duality group $SL(2)$. However, they are completely neutral under $SU(3)\times U(1)$, they are in the representation $\mathbf{1}_0$.

The very last bosonic fluctuations are the $\mathbf{6}_c$ tensors denoted $B_{\mu \nu} ^{i\alpha}$. In the black hole background, their equations of motion \eqref{eq: tensor eom} simplify to:
\begin{equation}
\label{eq: Bfield eom}
    \epsilon^{\mu \nu \rho \sigma \tau} D_\rho B_{\sigma \tau} ^{i\pm}+ i B^{i\pm}_{\mu \nu} =0~,
\end{equation}
where 
\begin{equation}
B^{\mu \nu i \pm} = \frac{1}{2}(B^{\mu \nu i 1}+ \Omega^{ij}B^{\mu \nu j2})\pm \frac{i}{2} (B^{\mu \nu i2}-\Omega^{ij}B^{\mu \nu j1})~,
\end{equation}
is the complexified tensor that diagonalizes the equations of motion. It has covariant derivative:
\begin{equation}
\label{eq: Bfield charge}
    D_\rho B_{\sigma \tau} ^{i\pm} = \partial_\rho B_{\sigma \tau} ^{i\pm} \mp g A_\mu B^{j\pm} _{\sigma \tau}~. 
\end{equation}
Basically, there are $12$ independent tensor fields, each a doublet under the $SL(2)$ duality group, and equally many with each sign of the minimal coupling to the background gauge field. This confirms the $\mathbf{3}_2 + \mathbf{\bar{3}}_{-2}$ charge assignment under $SU(3)\times U(1)$ identified in section \ref{sec:ssb}.

\subsection{Summary: Bosonic Fluctuations}
\label{subsec:fluceom}
As an overview of the equations of motion we tabulate some properties of the $42$ scalar and $26$ vector fluctuating fields. In our normailzation, the $U(1)$ assignment is twice the magnitude of the charge $e$ that appears in the minimal coupling, ie. the gauge covariant derivative $D_\mu = \partial_\mu + egA_\mu$.

\begin{table}[h]
    \centering
    \begin{tabular}{ccccccc}
          & Mixing & Charge & \# components & $SU(3)\times U(1)$ &  Pauli $p$\\ 
          \hline
         $t^+$ & $\times$ & $\pm 2$ & 12 & $\bar{\mathbf{6}}_4 + \mathbf{6}_{-4}$  & $+2$\\
         \hline
         $t^-$ &  $\checkmark (a^-)$  & 0 & 8 & $\mathbf{8}_0$  & $-2$ \\
         \hline
         $\Lambda_\alpha ^{\ \beta}$ & $\times$ & 0 & 1 (complex) & $\mathbf{1}_0$&   0 \\ 
         \hline 
         $\varphi_{ijk\alpha} ^{\mathbf{1}}$  & $\times$ & $\pm 3$ & 1 (complex) & $\mathbf{1}_{-6}+\mathbf{\bar{1}}_6$ &0 \\ 
         \hline
         $\varphi_{ijk\alpha} ^{\mathbf{3}}$  & $\times$ & $\pm 1$ & 3 (complex) & $\mathbf{3}_{-2}+\mathbf{\bar{3}}_2$ &  0\\ 
         \hline
         $\varphi_{ijk\alpha} ^{\mathbf{6}}$  & $\times$ & $\pm 1$ & 6 (complex) & $\mathbf{6}_{2}+\mathbf{\bar{6}}_{-2}$ &  +4 \\ 
         \hline
    \end{tabular}
    \caption{The $\mathbf{20}'+ \mathbf{10}_c+\mathbf{1}_c$ scalar fields as perturbations of the Kerr-Newman-AdS black hole, and some of their properties. The electric charge $e$ is defined with the normalization $D_\mu = \partial_\mu + ie A_\mu$ and the Pauli coupling via the effective mass $m^2_{\rm eff} = -\frac{1}{2}p F_{\mu \nu} F^{\mu \nu}$.}
    \label{tab:scallabel}
\end{table}

We introduce reference to Pauli-like couplings with the shorthand ``Pauli". These are couplings directly to the background field strength $F_{\mu \nu}$, rather than the gauge field $A_\mu$. For scalars, the Pauli-couplings are quadratic in the field so, when the field is constant, they act like a mass term that we normalize so $m^2_{\rm eff} = - \frac{1}{2}p F_{\mu \nu} F^{\mu \nu}$. For vectors, we bring attention to $F_{\mu\nu}$-couplings by a check-mark. 

For some fields, namely $t_-$ and $a_-$, there is not an independent equation of motion for scalars and another for the vector, the fields couple to one another. This feature is highlighted in the columns denoted ``Mixing". 

\begin{table}[h]
    \centering
    \begin{tabular}{cccccc}
          & Mixing & Charge & \# components & $SU(3)\times U(1)$ &  Pauli\\ 
          \hline
                  $a^+$ & $\times$ & $\pm 2$ & 6 & $\mathbf{3}_4 + \bar{\mathbf{3}}_{-4}$ &  \checkmark \\
         \hline
         $a^-$ &  $\checkmark (t^-)$ & 0 & 8 & $\mathbf{8}_0$ & \checkmark \\
         \hline 
         $B_{\mu \nu} ^{i\pm}$ & $\times$ & $\mp$ 1 & 6 (complex) & $2(\mathbf{3}_2 + \mathbf{\bar{3}}_{-2})$  & 0\\
         \hline
    \end{tabular}
    \caption{The vector fields as perturbations of the Kerr-Newman-AdS black hole, and some of their properties.}
    \label{tab:veclabel}
\end{table}

\section{The Near Horizon Geometry}
\label{sec:nearhg}
In this section, we work out details of the Kerr-Newman-AdS black hole for the extremal case, with focus on the near horizon AdS$_2$ region. This will provide a tractable setting where the equations of motion for the fluctuation can be solved, as we do in the next section. 

\subsection{The Extremal Limit of Kerr-Newman-AdS Black Holes}
The KNAdS geometry \eqref{eq: KNAdS geom} is quite unwieldy. To make progress, we introduce the invariant one-forms $\sigma_i$: 
\begin{align}
    \label{eq: sigma1 def}
    \sigma_1 &= -\sin \psi d\theta + \cos\psi \sin \theta d\phi ~, \\ 
    \label{eq: sigma2 def}
    \sigma_2 &= \cos \psi d\theta + \sin\psi \sin \theta d\phi ~, \\ 
    \label{eq: sigma3 def}
    \sigma_3 &= d\psi + \cos\theta d\phi ~,
\end{align}
that satisfy $d\sigma_i = \epsilon_{ijk} \sigma_j \sigma_k$. Reorganization then gives the line element\footnote{In this section we take $g=\ell_5^{-1}=1$.}: 
\begin{equation}
    \label{eq: KNAdS geom sigmas} 
    ds^2 = -\frac{r^2}{(r^2+a^2)^2(1-a^2)} \Delta_r e^{U_2 - U_1} dt^2 + \frac{(r^2+a^2)dr^2}{\Delta_r} + e^{-U_1} (\sigma_1 ^2 + \sigma_2 ^2) + e^{-U_2} (\sigma_3 + \omega dt)^2 ~, 
\end{equation}
where the scalar functions $U_1$, $U_2$, and $\omega$ are defined through: 
\begin{align}
    \label{eq: U1 def} 
    e^{-U_1} &= \frac{r^2 + a^2}{4(1-a^2)} ~, \\ 
    \label{eq: U2 def} 
    e^{-U_2} &= \frac{r^2+a^2}{4(1-a^2)}\left(1+\frac{2(m+q)a^2}{(1-a^2)(r^2 +a^2)^2} - \frac{a^2 q^2}{(1-a^2)(r^2+a^2)^3}\right) ~, \\
    \label{eq: geom omega def} 
    \omega &= \frac{a}{2(1-a^2)} \left((r^2+a^2) -\frac{2m+q}{r^2+a^2} + \frac{q^2}{(r^2 +a^2)^2}\right)e^{U_2} -2a ~.
\end{align}
The gauge field \eqref{eq: KNAdS A pot} supporting the solution becomes
\begin{equation}
\label{eqn:calAnew}
A
 = \frac{q}{(1-a^2)(r^2+a^2)} (dt - \frac{1}{2}a \sigma_3) ~.
\end{equation}

We will focus on the extremal limit where the temperature $T$ given in \eqref{eqn:Tdef} vanishes and so the parameters of the solution are related as
\begin{equation}
\label{eqn:qextdef}
q = r^2_+ \sqrt{2r^2_+ + 1 + 2a^2}-a^2~.
\end{equation}
The coordinate position of the horizon $r_+$, is given by the vanishing of $\Delta_r$, as in \eqref{eqn:horeq}. 
This equation can be reorganized to give the parameter $m$:
\begin{equation}
\label{eqn:mextdef}
2m = (r^2_++a^2)(3r^2_++2 + a^2)~.
\end{equation}
In the extremal limit, the physical charges (\ref{eqn:Qdef}-\ref{eqn:Jdef}) become: 
\begin{align}
\label{eqn:Qext}
Q & = \frac{\pi}{4G_5}\frac{1}{(1-a^2)^2}\left(r^2_+ \sqrt{2r^2_+ + 1 + 2a^2}-a^2\right)~,\\
\label{eqn:Jext}
J & = \frac{\pi}{4G_5}\frac{a}{(1-a^2)^3}
\left( 3r^4_+ + 2r^2_+ (1+ 2a^2) + a^2 +r_+ ^2(1+a ^2)\sqrt{2r^2_++1+2a^2}\right)~.
\end{align}
For extremal black holes, the variables $(r_+,a)$ parametrize the physical charges $(Q,J)$ through these equations. Equivalently, the parameters $(r_+,a)$ correspond to the thermodynamic potentials (\ref{eqn:Omdef}-\ref{eqn:Phidef}) through 
\begin{align}
\label{eqn:phidef}
\Phi & = 
\frac{-a^2 + r^2_+ \sqrt{2r^2_+ + 2a^2+1}}{r^2_+ + 2a^2 + a^2\sqrt{2r^2_+ + 2a^2+1}}~,
\\
\label{eqn:omdef}
\Omega & = a 
\frac{r^2_++a^2 +1 +\sqrt{2r^2_++1 + 2a^2}}{r^2_++2a^2 + a^2\sqrt{2r^2_+ + 1 + 2a^2}}~.
\end{align}
We also record the entropy \eqref{eqn:Sdef} specialized to the extremal limit: 
\begin{equation}
S  = \frac{\pi^2}{2G_5} \frac{r_+ (r^2_+ +2a^2 + a^2 \sqrt{2r^2_+ +2a^2+1})}{(1-a^2)^2} ~.
\end{equation}

We assume $Q\geq 0$ without loss of generality, corresponding to parameters satisfying 
\begin{equation}
\label{eqn:rp range}
r^2_+ \geq \frac{1}{4}(\sqrt{1+8a^2}-1)~.
\end{equation}
Finiteness of charges restrict parameters to the range $0\leq a\leq 1$. The family of extremal black holes interpolate between extremal Kerr that rotates, but has no charge, and extremal Reissner-Nordstr\"{o}m that has charge but is spherically symmetric. In between, when the charges are balanced according to the constraint
\begin{equation}
\label{eq:nonlinear charge const}
\left(Q^3 + \frac{\pi}{4G_5}J^2\right)-\left(3Q + \frac{\pi}{4G_5}\right)\left(3Q^2 - \frac{\pi}{2G_5}J \right)=0~,
\end{equation}
is satisfied, the black hole is supersymmetric. That is precisely when the mass excess 
$$
M- 3Q -J = \frac{\pi}{4G_5}\frac{3-a}{(1-a)^2(1+a)^3}
\left( \frac{1}{2}(r^2_++a^2) (3r^2_++2 + a^2) - 
(1+2a)(r^2_+ \sqrt{2r^2_+ + 1 + 2a^2}-a^2)\right)
\geq 0~,
$$
saturates the BPS bound, given by the inequality on the right hand side. That happens when the parameter $r_+$ is given by
\begin{equation}
\label{eqn:rstardef}
r^2_* = a(2+a)~.
\end{equation}

\begin{figure}[H]
    \centering
    \includegraphics[scale=0.8]{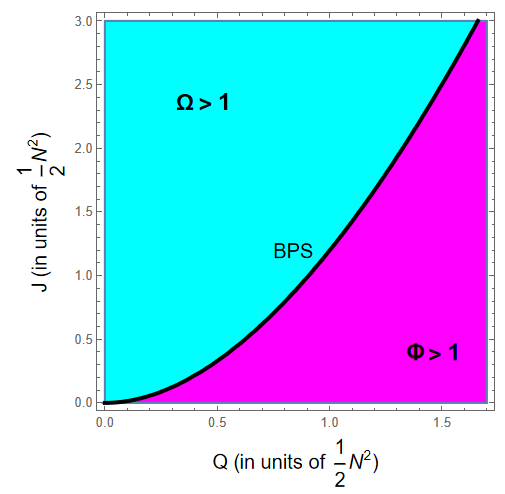}
    \caption{Parameter space of $Q$ and $J$ relative to the (BPS) nonlinear charge constraint \eqref{eq:nonlinear charge const}, in $\text{AdS}_5/\text{CFT}_4$ units of $\tfrac{1}{2}N^2 = \frac{\pi}{4G_5}$.}
    \label{fig:QJ plot}
\end{figure}

It is interesting to study the extremal AdS black holes as they interpolate from only angular momentum (Kerr), through BPS, to only charge (Reissner-Nordstr\"{o}m). On the ``mostly charged" side of the BPS line, the electric potential $\Phi>1$, while on the ``mostly angular momentum side", the rotational velocity $\Omega>1$. These regions meet at the BPS line where $\Phi=\Omega=1$ (Figure
\ref{fig:QJ plot}). 
The extremal black holes depend on one less parameter than the generic KN-AdS black holes, but their geometry is not greatly simplified. For example, all black holes with ``mostly rotation" feature a non-trivial ergoregion. It is located where $g_{tt}=0$, equivalent to
$$
\frac{r^2}{(1-a^2)(r^2+a^2)^2}\Delta_r e^{U_2-U_1} = e^{-U_2}\omega^2~.
$$
Therefore, in order to render the equations manageable, we simplify further. In the next subsection we specialize to the near horizon region. 

\subsection{The Near-horizon Limit of Extremal Kerr-Newman-AdS Black Holes}

For extremal black holes, the near horizon geometry necessarily includes an AdS$_2$ component, but it can be non-trivially fibered over the AdS$_2$. We write a generic near horizon geometry with $SU(2)$ symmetry as: 
\begin{align} \label{eq: nh ext KN geom}
    ds^2 &= -A^2 (r-r_+ )^2 dt^2 + \frac{B^2}{(r-r_+)^2}dr^2 + \frac{1}{4}C^2 (\sigma_1 ^2 + \sigma_2 ^2) \nonumber \\
    &\quad + \frac{1}{4}D^2 \left(\sigma_3 + E(r-r_+) dt\right)^2 ~. 
\end{align}
That $g_{tt}$ and $g_{rr}^{-1}$ have double poles at the horizon is characteristic of AdS$_2$. The angular form $\sigma_3$ in the horizon geometry \eqref{eq: nh ext KN geom} is defined as co-moving, so the rotation term proportional to $dt$ vanishes at the horizon. This horizon convention differs from the asymptotic notation in \eqref{eq: KNAdS geom sigmas} as
\begin{equation}
\label{eqn:horvsas}
\sigma_{3{\rm as}} = \sigma_{3{\rm hor}}-\omega\Big|_{r=r_+}dt
= \sigma_{3{\rm hor}}+2\Omega dt~,
\end{equation}
where $\Omega$ is given in \eqref{eqn:omdef}~. The coefficients $\{A,B,C,D,E\}$ are numbers, in that they are independent of spacetime position. By expanding the explicit geometry \eqref{eq: KNAdS geom sigmas} near the horizon $r=r_+$, we find the two parameter family of near horizon geometries:
\begin{equation} \label{eq: ABCDE def}
\begin{split}
        A &= \frac{2\sqrt{(r_+ ^2 + a^2)(3r_+ ^2 +2a^2 +1)}}{r_+ ^2 +2a^2 + a^2 \sqrt{2r_+ ^2 +1 +2a^2}} ~,\\
        B & = \sqrt{\frac{r_+ ^2 +a^2}{4(3r_+ ^2 +2a^2 +1)}} ~, \\ 
    C &= \sqrt{\frac{r_+ ^2 +a^2}{1-a^2}} ~,
    \\ 
 D &= \frac{r_+}{(1-a^2)(r_+ ^2 +a^2) }
 \left(r_+ ^2 +2a^2 + a^2 \sqrt{2r_+ ^2 +1 +2a^2}\right) ~,
    \\
    E &= \frac{4a(1-a^2)}{r_+} \frac{r_+ ^2 +(2r_+ ^2 +a^2)\sqrt{2r_+ ^2 +1 +2a^2}}{\left(r_+^2 +2a^2 + a^2 \sqrt{2r_+ ^2 + 1 +2a^2}\right)^2} ~.
\end{split}
\end{equation}
We denote the near horizon gauge field ${\cal A}$, not to be confused with the parameter $A$ in the near horizon geometry \eqref{eq: nh ext KN geom}. 
We pick a gauge so 
\begin{equation} \label{eq: At=0 hor}
    \mathcal{A}_t \vert_{r=r_+} = 0 ~. 
\end{equation}
Generally, this requires shifting the potential $\mathcal{A}$ by a constant, which is a pure gauge term generated by a gauge function linear in $t$. With this choice, we write a generic near horizon gauge field as 
\begin{equation}
\label{eq: NH A pot generic}
    \mathcal{A} =  -F (r-r_+)dt - G \sigma_3 ~.
\end{equation}
where the coefficients $F$ and $G$ are numbers. This gives the field strength:
\begin{equation} \label{eq: KN F def}
    \mathcal{F} =d\mathcal{A} = -F dr dt -G \sigma_1 \sigma_2 ~. 
\end{equation}
Note that we now refer to the background field strength as $\mathcal{F}$, with $F$ to be reserved for its electric component. 

The gauge potential $\mathcal{A}$ \eqref{eqn:calAnew} in the explicit solution applies the asymptotic convention for the angular form $\sigma_3$ that must be shifted to the co-moving horizon frame through \eqref{eqn:horvsas}. Further, we must add a pure gauge term to $\mathcal{A}_t$, in order to satisfy \eqref{eq: At=0 hor}. We then have 
\begin{align} \label{eq: KN pot shifted}
    \mathcal{A} &= \frac{q}{(1-a^2)}\left(1-a\Omega\right)\left(\frac{1}{r^2 +a^2} -\frac{1}{r_+ ^2 +a^2}\right) dt -\frac{qa}{2(1-a^2)(r^2 +a^2)}\sigma_3 ~, 
\end{align}
which in the near-horizon region simplifies to:
\begin{equation}
\label{eq: NH A pot def}
    \mathcal{A} = -\frac{2qr_+}{(r_+ ^2 +a^2)(r_+ ^2 +2a^2 +a^2\sqrt{2r^2_++1+2a^2})}(r-r_+)dt -\frac{qa}{2(1-a^2)(r_+ ^2 +a^2)}\sigma_3  ~.
\end{equation}
This takes the form \eqref{eq: NH A pot generic} with: 
\begin{equation} \label{eq: FG def}
\begin{split}
F &=\frac{2\left(r_+ ^2 \sqrt{1+2a^2 +2r_+ ^2}-a^2 \right)r_+}{\left(r_+ ^2 +a^2)(r_+ ^2 +2a^2 +a^2\sqrt{2r^2_++1+2a^2}\right)}~,  \\
G &=\frac{\left(r_+ ^2 \sqrt{1+2a^2 +2r_+ ^2}-a^2 \right)a}{2(1-a^2)(r^2 +a^2)}~.
\end{split}
\end{equation}

We will regularly need the Lorentz invariant square of $\mathcal{A}$ given by the contraction: 
\begin{equation}
\begin{split}
\label{eqn:Asqdef}
    \mathcal{A} \wedge \star \mathcal{A} &= \mathcal{A}_\mu \mathcal{A}^\mu \star 1 = \left(\frac{4G^2}{D^2} - \frac{(F-EG)^2}{A^2}\right) \star 1  \\
    &= \frac{\left(a^2 - r_+ ^2 \sqrt{1+2a^2 +2r_+ ^2}\right)^2 \left(-r_+ ^2 +a^2 (1+3r_+ ^2 +3a^2+2\sqrt{1+2a^2 +2r_+ ^2})\right)}{(r_+ ^2 +a^2)(1+2a^2 +3r_+ ^2)\left(r_+ ^2 +a^2 (2+\sqrt{1+2a^2 + 2r_+ ^2})\right)^2} \star 1 ~, 
\end{split}
\end{equation}
and the Lorentz invariant square of $\mathcal{F}$ given through:
\begin{equation}
\begin{split}
    \label{eqn:Fsqdef}
    \mathcal{F}\wedge \star \mathcal{F} = \frac{1}{2}\mathcal{F}_{\mu \nu} \mathcal{F}^{\mu \nu} \star 1 &= \left(\frac{16G^2}{C^4}-\frac{F^2}{A^2 B^2}\right)\star 1 \\ 
    &= \frac{4(a^2 -r_+ ^2)\left(a^2 - r_+ ^2 \sqrt{1+2a^2+2r_+ ^2}\right)^2}{(r_+ ^2 +a^2)^4} \star 1 ~. 
\end{split}
\end{equation}
We have derived explicit relations for the parameters $\{A,B,C,D,E,F,G\}$ by taking a limit of the KNAdS black hole solution. It is instructive to show that {\it all} near horizon solutions arise in this way. To do so, we consider the general ans\"{a}tze for the geometry \eqref{eq: ABCDE def} and the gauge field \eqref{eq: KN F def}. The Maxwell-Chern-Simons equation of motion $d\star \mathcal{F} + \mathcal{F}\wedge \mathcal{F}=0$ imposes 
\begin{equation} \label{eq: nh params rel maxwell}
    E = \frac{16AB(ABD+C^2 F)}{C^4  D F} G ~, 
\end{equation}
after some algebra. 
This relates $E$ and $G$, the two coefficients that are odd under time-reversal. 
In the near horizon geometry, Einstein's field equations: 
\begin{equation}
    R_{\mu \nu} +4g_{\mu \nu} -\left(\frac{3}{2}\mathcal{F}_{\mu \lambda} \mathcal{F}_{\nu} ^{\ \lambda} -\frac{1}{4}g_{\mu \nu} \mathcal{F}_{\lambda \rho} \mathcal{F}^{\lambda \rho} \right) = 0 ~, 
\end{equation}
reduce to three independent conditions, from the AdS$_2$, $S^2$, and $S^1$ components of Einstein equations. After some effort we find:
\begin{align}
\label{eq: nh params rel einstein1}
    B &= \frac{C}{2\sqrt{1+3C^2}} ~, \\ 
\label{eq: nh params rel einstein2}
    E &= \frac{1}{CD}\sqrt{\frac{2A^2 D^2}{1+3C^2} + 8A^2 C^2-6C^2 F^2} ~, \\ 
\label{eq: nh params rel einstein3}
    G &= \frac{1}{2\sqrt{2}A}\sqrt{A^2 (2C^2 + 2C^4 -D^2 ) - C^2 (1+3C^2)F^2} ~. 
\end{align}

The parametrization of a near horizon black hole solution in terms of the $7$ independent coefficients $\{A,B,C,D,E,F,G\}$ has one continuous redundancy. Namely, rescaling of the time-coordinate $t$ is equivalent to simultaneous scaling of $A$, $E$, and $F$. Therefore, the four conditions \eqref{eq: nh params rel maxwell} and (\ref{eq: nh params rel einstein1}-\ref{eq: nh params rel einstein3}) identify a two parameter family of solutions. 
The solutions \eqref{eq: ABCDE def} and \eqref{eq: NH A pot def} that we computed from the near horizon limit of the KNAdS black hole do in fact depend on precisely two independent parameters, denoted $\{a,r_+\}$, and they solve
\eqref{eq: nh params rel maxwell} and (\ref{eq: nh params rel einstein1}-\ref{eq: nh params rel einstein3}). Therefore, this must be the most general solution. This result also serves as a useful validation of our algebra. 


\subsection{The Scalar Wave Equation in $\text{AdS}_2$}
In the near-horizon region, nearly all the equations of motion will ultimately reduce to a massive scalar field that propagates in $\text{AdS}_2$ spacetime. The two-dimensional line element, in a co-rotating frame where $d\psi$ in $\sigma_3$ \eqref{eq: sigma3 def} is shifted by $E (r-r_+) dt$, can be summarized as:
\begin{equation}
    ds_2 ^2 = -A^2 (r-r_+)^2 dt^2 + \frac{B^2}{(r-r_+)^2} dr^2 ~. 
\end{equation}
The coordinate transformation:
\begin{equation}
\label{eq: poincare ads2 coord transf}
    z = \frac{1}{r-r_+} \ , \ \tau =\frac{B}{A} t ~,  
\end{equation}
yields $\text{AdS}_2$ in the familiar Poincar\'{e} coordinates:
\begin{equation}
\label{eq: ds2 poincare form}
    ds_2 ^2 = \frac{B^2}{z^2} \left(-d\tau^2 + dz^2\right) ~. 
\end{equation}
In this form it is manifest that the $\text{AdS}_2$ scale is $\ell_2 = B$. 

In the near horizon region, the massive Klein-Gordon equation for a scalar field $\Phi$ with time-dependence $e^{i\omega t}$ is:
\begin{align}
    (\Box -m^2)\Phi &= 
    \left( \tfrac{1}{\sqrt{-g}} \partial_\mu \left(\sqrt{-g} \partial^\mu\right)-m^2\right)\Phi \cr
    & = \left(\frac{1}{B^2} \partial_r ((r-r_+)^2 \partial_r ) +\frac{\omega^2}{A^2 (r-r_+)^2} -m^2\right) \Phi =0~. 
\end{align}
The Poincar\'{e} coordinates \eqref{eq: poincare ads2 coord transf} give a standard Schr\"{o}dinger-like wave equation: 
$$
\left( - \frac{\partial^2}{\partial z^2} + \frac{m^2 B^2}{z^2} \right) \Phi =\tfrac{B^2}{A^2}\omega^2 \Phi ~.
$$
The $\frac{1}{z^2}$-potential is typically repulsive, and all solutions are scattering states. However, for negative coefficient the potential is attractive and, when the coefficient is sufficiently negative, it supports bound states. That corresponds to imaginary $\omega$, so such solutions have exponential time-dependence and are unstable. Stability therefore imposes a lower bound on the mass of the scalar, in units of the near-horizon $\text{AdS}_2$ length scale $\ell_2 = B$: 
\begin{equation}
\label{eqn:BFbound}
    m^2 \ell^2_2\geq -\frac{1}{4} ~. 
\end{equation}
This is the standard Breitenlohner-Freedman bound \cite{breitenlohner1982positive}. It is equivalent to the condition that the conformal dimension in AdS$_2$
$$
\Delta_0 = \frac{1}{2} + \sqrt{\frac{1}{4} + m^2\ell^2_2}~, 
$$
is real.

The scalar fields of ${\cal N}=8$ gauged supergravity in $D=5$ are at best massless, but most are ``tachyonic", they all satisfy $m^2\ell^2_5\leq 0$. However, they are stable, because they satisfy the 5D Breitenlohner-Freedman bound $m^2 \ell^2_5\geq -4$. The analogous bound in 2D \eqref{eqn:BFbound} appears to be more strict, by the relative factor $4$ vs $\frac{1}{4}$. However, a precise comparison depends critically on the $\text{AdS}_2$ length scale given in \eqref{eq: ABCDE def} as 
\begin{equation}
\label{eqn:AdS2scale}
\ell_2 = B = \sqrt{\frac{r^2_++a^2}{4(3r^2_++2a^2+1)}}~, 
\end{equation}
in units where $\ell_5=1$~. 

For example, a minimally coupled scalar with $m^2=-4$ in AdS$_5$ is stable in the near-horizon AdS$_2$ region if the black hole charges, parametrized by $(r_+,a)$, are such that $\ell_2\leq \frac{1}{4}$. This condition can go either way. Specifically, each of the Kerr, BPS, and Reissner-Nordstr\"{o}m families of black holes permit the range $\ell_2\in (0,\tfrac{1}{2\sqrt{3}})$ so, for any of these slices of parameters, a field with $m^2=-4$ may or may not be stable. On the other hand, by the same criterion, a scalar with $m^2=-3$ in AdS$_5$ has $m^2\ell^2_2\geq -\frac{1}{4}$ so it is always stable.

This cannot be the complete story. In particular, BPS black holes in supergravity must be stable. The catch is that, as we develop in this article, fluctuating fields couple non-trivially to the black hole background. Our analysis in the next section will recast the non-minimal couplings as additional contributions to the effective mass, and then apply the bound  \eqref{eqn:BFbound} on the aggregate.  

We will also study generalizations of scalars to 
$p$-form fields $\Phi_p = \frac{1}{p!}\Phi_{\mu_1 ... \mu_p} dx^{\mu_1} ... dx^{\mu_p}$ with the schematic Lagrangian: 
\begin{equation}
    \mathcal{L} = d\Phi_p \wedge \star d\Phi - m^2 \Phi_p \wedge \star \Phi     ~~\leftrightarrow~~ d\star d\Phi -(-1)^p m^2 \star \Phi_p = 0~.
\end{equation}
For genuine forms $p\neq 0$ the massive equations of motion imply the consistency relation:
\begin{equation} 
\label{eq: pform consistency}    d\star \Phi_p = 0 ~~,~~~(m^2 \neq 0)~. \end{equation}
In $D=5$ spacetime dimensions, a $p$-form field has ${D-1 \choose p} ={ 4 \choose p}$ on-shell degrees of freedom, so a $1$-form vector has $4$ d.o.f's and a $2$-form massive tensor $6$ d.o.f's.
In the near-horizon $\text{AdS}_2$ region, we can interpret the free massive $p$-form  equation of motion as ${ 4 \choose p}$ effective scalar fields.


\section{Fluctuating Supergravity Fields in the Near Horizon Region}
\label{sec:nearh}

In this section we study the equations of motion for fluctuations in the near horizon region of the KNAdS black hole in ${\cal N}=8$ supergravity. The main focus is the interplay between non-minimal couplings and the BF stability bound. 

\subsection{The Effective Mass in the Near Horizon Region} 
We will find that, in the near horizon region, most field equations are ultimately equivalent to scalar fields, albeit with a few complications. First of all, many of the fields are subject to the familiar minimal coupling of a scalar field to a background gauge field
\begin{equation}
\begin{split}
    D\phi &= d\phi  + i e \mathcal{A} \phi ~. 
\end{split}
\end{equation}
Additionally, we must account for the ``Pauli"-coupling directly between the field and the background field strength $F_{\mu\nu}$. Altogether, we present the 5D Lagrangian for ``simple" scalars as: 
\begin{equation}
    \mathcal{L} = -\frac{1}{2}d\phi\wedge d\phi^* -\frac{1}{2}\left( m^2 \star 1 + e^2 \mathcal{A}\wedge \star \mathcal{A}-p \mathcal{F}\wedge \star \mathcal{F}\right)\phi\phi^*  ~. 
\end{equation}
The gauge potentials and the fields strengths are constant in the near horizon region, so there the equations of motion reduce to the free, massive, Klein-Gordon equation for a complex scalars with effective mass: 
\begin{equation}
\label{eqn:effmass}
m^2_{\rm eff} = m^2  + e^2 \mathcal{A}_\mu\mathcal{A}^\mu 
- \frac{1}{2}p  \mathcal{F}_{\mu\nu}\mathcal{F}^{\mu\nu}~. 
\end{equation}
The mass-term $m^2$ is the mass that is familiar from fields in the AdS$_5$ vacuum. The remaining terms depend on the background electromagnetic fields through the Lorentz invariant fields combinations given in \eqref{eqn:Asqdef} and \eqref{eqn:Fsqdef}, respectively. In this section we confront the effective mass \eqref{eqn:effmass} with the BF stability criterion for scalars  \eqref{eqn:BFbound}.

The additional terms $\mathcal{A}_\mu\mathcal{A}^\mu$ and $\mathcal{F}_{\mu\nu}\mathcal{F}^{\mu\nu}$ vanish for the AdS-Kerr black hole, which has no electromagnetic fields. Therefore, the intuition from pure AdS$_5$ provides good guidance near this limit. In the complementary limit, the Reissner-Nordstr\"{o}m-AdS black hole that does not rotate, the contribution of the minimal coupling term $+e^2  \mathcal{A}_\mu\mathcal{A}^\mu $ to \eqref{eqn:effmass} is negative in our conventions, and it may result in an instability that drives the vacuum superconducting in the near horizon region \cite{hartnoll2008holographic, basu2010small, bhattacharyya2010small, horowitz2011introduction}. On the other hand, the prevailing sign of the Pauli coupling is $p\geq 0$ so, for the Reissner-Nordstr\"{o}m-AdS black hole, it tends to compensate the destabilizing $\mathcal{A}_\mu\mathcal{A}^\mu$. 

\begin{figure}[H]
    \centering
    \includegraphics[scale=0.6]{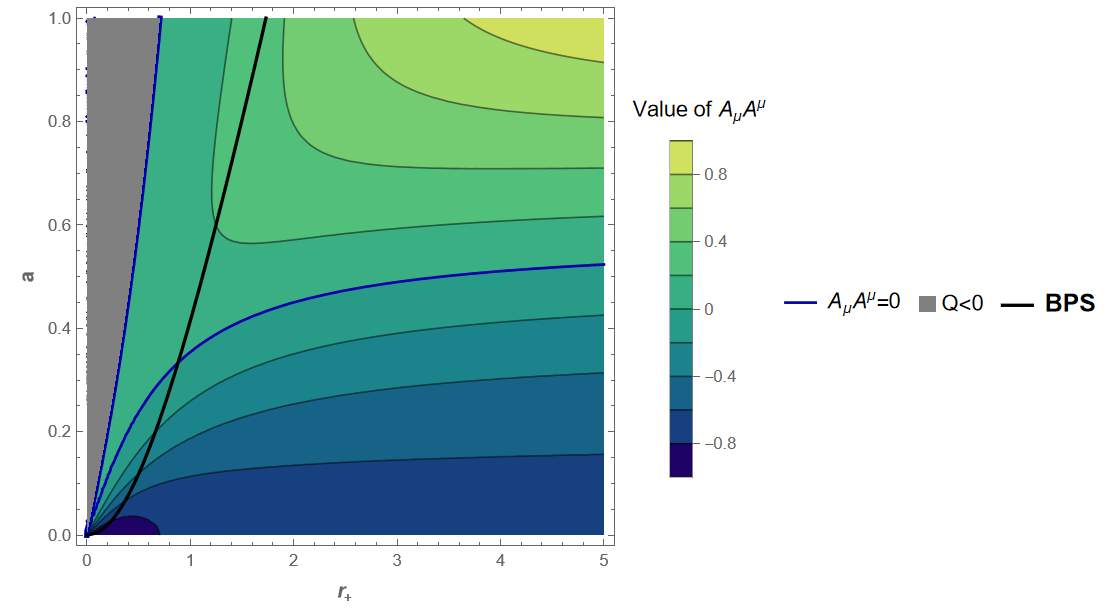}
    \caption{The Lorentz invariant version of the squared potential $\mathcal{A}_\mu \mathcal{A}^\mu$ in the near horizon region, as function of parameters $(r_+, a)$, excluding the $Q<0$ unphysical region (grey). The BPS line \eqref{eqn:rstardef} is in black. The invariant is negative and relatively large near the horizontal axis, corresponding to a significant electric field in the RNAdS near horizon region. As rotation increases, the invariant changes sign at the blue curve that extends from the origin to large $r_+$ values. Near the KNAdS limit, at the boundary to the grey region that hugs the vertical axis, the invariant is positive and relatively small, corresponding to a modest magnetic potential. The negative values that prevail when charge dominates over rotation contribute towards an instability, except for neutral fields where this contribution vanishes.}
    \label{fig:A2 sign}
\end{figure}

\begin{figure}[H]
    \centering
    \includegraphics[scale=0.6]{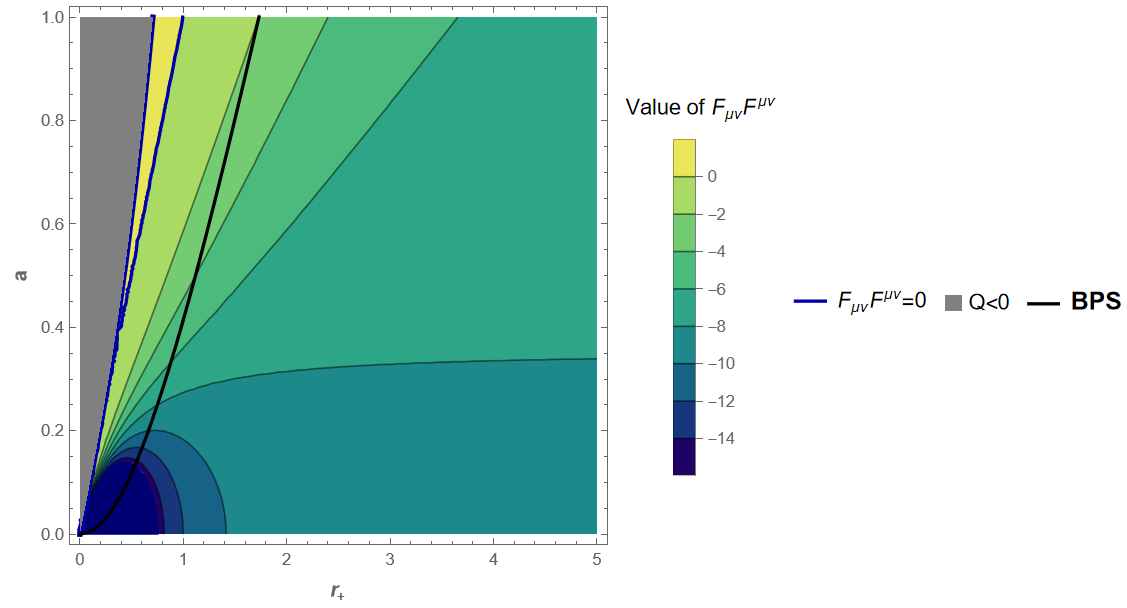}
    \caption{The Lorentz invariant version of the squared field strength $\mathcal{F}_{\mu\nu} \mathcal{F}^{\mu \nu}$ as a function of $(r_+, a)$, excluding the $Q<0$ unphysical region  (grey). The invariant is negative over nearly the entire physical parameter space, corresponding to an electric field, but there is a small sliver between the sign-change locus $(r_+ =a)$ and the Kerr $(Q=0)$ boundary.} 
    \label{fig:F2 sign}
\end{figure}

For reference, in Figures \ref{fig:A2 sign} and \ref{fig:F2 sign}, we plot the Lorentz invariants $\mathcal{A}_\mu \mathcal{A}^\mu$ and $\mathcal{F}_{\mu\nu} \mathcal{F}^{\mu \nu}$. We parametrize the physical region of extremal KNAdS backgrounds by $(r_+,a)$, with $0\leq a \leq 1$ and $r_+$ satisfies \eqref{eqn:rp range}, as required by the condition $Q\geq 0$ given in \eqref{eqn:qextdef}. On our plots, RNAdS is on the horizontal axis where $a=0$. Qualitatively, the vertical axis is a proxy for rotation, but more precisely the neutral KerrAdS limit is at the value where $r_+$ is minimal. The plots show that $\mathcal{A}_\mu \mathcal{A}^\mu<0$ and $\mathcal{F}_{\mu\nu} \mathcal{F}^{\mu \nu}<0$ near the RN-AdS limit, as expected for purely electric fields. On the other hand, in some regions of parameter space these invariants turn positive, associated with magnetic fields. 
In the section, we illustrate the balance between these effects using the probes available in supergravity.

\subsection{The Simple Scalars} 

We first study the near horizon fluctuations of the simplest scalars, i.e. the scalars that do not mix with vector fields.
For these, the behavior depends on the couplings $e$ and $p$. They were discussed for various fields throughout section \ref{sec:fluc} and tabulated in Table \ref{tab:scallabel}. Nearly all scalar fields are susceptible to several of the terms in \eqref{eqn:effmass}, and so their behavior depend on an interplay between several contributions. The only exceptions are the two fields denoted $\Lambda_\alpha ^\beta$. They have vanishing $\text{AdS}_5$ mass $m=0$ \eqref{eq: Lambda eom}, and also vanishing charges $e=0$. Thus $\Lambda_\alpha ^\beta$ are minimally coupled massless scalar fields that are often studied as the simplest spectator fields in a black hole background. It is satisfying that such probes are realized in ${\cal N}=8$ supergravity.

For the non-trivial scalars, we first consider $t^+$. These $12$ fields are a subset of the ${\bf 20}'$ scalars that, because they have $m^2=-4$, are precisely at the boundary of the BF-bound in the AdS$_5$ vacuum. According to Table \ref{tab:scal flucts}, the $t^+$ have electric charge $e=2$ and Pauli-coupling $p=2$. The intuition from the near horizon region of the RNAdS black hole is that electric charge $e$ destabilizes, but the Pauli coupling $p>0$ provides stabilization that may compensate. The complete formula for the effective mass in the near horizon region \eqref{eqn:effmass} is: 
\begin{equation}
\begin{split}
    m_{t^+} ^2&= -4+ 4\mathcal{A}_\mu \mathcal{A}^\mu -\mathcal{F}_{\mu \nu} \mathcal{F}^{\mu \nu} \\ 
    &= -4 +\frac{4 \left(a^2 \left(2 \sqrt{2 a^2+2 r_+^2+1}+3 a^2+3r^2_+ + 1\right)-r_+^2\right) \left(a^2-r_+^2 \sqrt{2
   a^2+2 r_+^2+1}\right)^2}{\left(a^2+r_+^2\right) \left(2 a^2+3 r_+^2+1\right) \left(a^2 \left(\sqrt{2 a^2+2
   r_+^2+1}+2\right)+r_+^2\right)^2}\\
   &\quad +\frac{8 \left(r_+^2-a^2\right) \left(a^2-r_+^2 \sqrt{2 a^2+2
   r_+^2+1}\right){}^2}{\left(a^2+r_+^2\right){}^4} ~. 
   \label{eqn:tplusmass}
\end{split}
\end{equation}


\begin{figure}
    \centering
    \begin{subfigure}[t]{0.5\textwidth}
        \centering
        \includegraphics[scale=0.5]{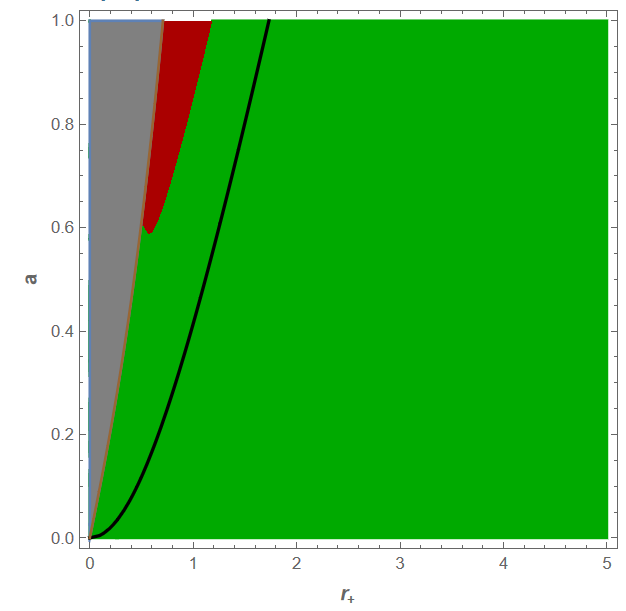}
        \caption{The actual parameter space for $t^+$ scalars. The 
        stability bound $m_{t^+} ^2 \ell_2 ^2 >-\tfrac{1}{4}$ (green) is satisfied, except for a small sliver near the Kerr boundary (dark red). These scalars are easily stable in the BPS background (black line).}
    \end{subfigure}%
    ~
    \begin{subfigure}[t]{0.5\textwidth}
        \centering
        \includegraphics[scale=0.5]{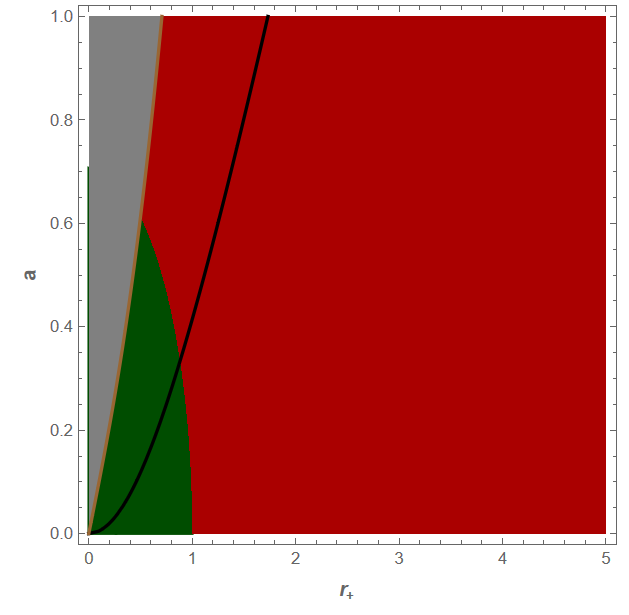}
        \caption{The parameter space for a hypothetical scalar with $m^2=-4$, but no charge or Pauli coupling ($e,p=0$). The BF stability criterion is violated (dark red) nearly everywhere, including on the BPS locus (black line).}
    \end{subfigure}
    \caption{The parameter space $(r_+,a)$ for the $t^+$ scalars. The horizontal axis corresponds to Reissner-Nordstr\"{o}m black holes (no rotation). The grey regions near the vertical axis are unphysical. Their inside boundary are the Kerr black hole (no charge). The black line through the middle is the BPS locus. The left panel is the actual parameter space. The right panel omits both the minimal and Pauli coupling. 
    \label{fig:t+ param space}
      }
\end{figure}

To study the near-horizon stability of the $t^+$ scalars by the BF-criterion \eqref{eqn:BFbound}, we must examine the sign of $m_{t^+} ^2 \ell_2 ^2 + \tfrac{1}{4}$ with $\ell_2 = B$ given in \eqref{eq: ABCDE def}. The expression for the mass \eqref{eqn:tplusmass}, and its analogue in units of $\ell_2$, are not illuminating as they stand, and we have not found better analytical expressions. We can show that, in general, $-2\leq m^2 \ell_2 ^2 \leq 2$, but this interval straggles the BF bound $m^2 \ell_2 ^2=-\frac{1}{4}$ that determines the qualitative behavior. The most useful analytical representation we have found is to record the range of possible effective masses in $\ell_2$ units along the Kerr, BPS and Reissner-Nordstr\"{o}m-curves: 
\begin{align}
\label{eq: tplus mass kerr}
    &\text{Kerr: } m^2 _{t^+} \ell_2 ^2 = \frac{1-\sqrt{1+8a^2}}{2\sqrt{1+8a^2}}\in (-\tfrac{1}{3},0) ~, \\ 
\label{eq: tplus mass BPS}
    &\text{BPS: } m^2 _{t^+} \ell_2 ^2 =  \frac{2+6a-4a^2}{(1+5a)^2} \in (\tfrac{1}{9},2) ~, \\ 
\label{eq: tplus mass RN}
    &\text{Reissner-Nordstr\"{o}m: } m^2 _{t^+} \ell_2 ^2 = \frac{2+8r_+ ^2 +7r_+ ^4}{(1+3r_+ ^2)^2} \in (\tfrac{7}{9}, 2) ~.
\end{align}
The ranges are $a\in (0,1)$ and $r_+\in(0,\infty)$. The Kerr-slice is stable when $a\to 0$ but, for larger $a$, it includes a range that is unstable by the criterion $m^2 _{t^+} \ell_2 ^2<-\frac{1}{4}$ \eqref{eqn:BFbound}. The couplings to the electric field do not affect Kerr, so this illustrates that the marginal condition $m^2 \ell_5^2=-4$ translates, in units of $\ell_2$, to stability for slowly rotating extremal black holes, but instability for the fastest ones. The analytical formulae suggest that, when black hole charge is turned on, the black holes quickly become ever more stable. 

The left panel in Figure \ref{fig:t+ param space} shows that indeed, the $t^+$ scalars satisfy the BF stability bound (lighter green) given by \eqref{eqn:BFbound} throughout most, but not all, of parameter space. The right panel in Figure \ref{fig:t+ param space} omits both the minimal coupling and Pauli-like terms. It shows that the scalar fields with $m^2 =-4$ familiar from AdS$_5$ holography are unstable (dark red) in nearly all extremal black hole backgrounds, before taking coupling to background gauge fields into account. In particular, such scalars would destabilize a portion of the BPS curve.

\subsection{The Pseudo-Scalars} 

The pseudo-scalars $\varphi_{ijk\alpha}$ are novel: we are not aware of any studies of their propagation in black hole backgrounds. That is unfortunate, because they do not appear at linear order so they do not source any of the commonly considered backgrounds. Therefore, they are natural as probes, also in settings that are more general than ours. 

The pseudo-scalars have mass $m^2\ell^2_5=-3$ in pure AdS$_5$, corresponding to conformal dimension $\Delta=3$. As such, they exhibit some propensity for instability, but they are not right on the BF-bound. On the other hand, they can have minimal couplings with very large charge, $\pm3$ in our units. Since that is at any rate before taking into account the Pauli couplings, and the distinction between $\ell_5$ and $\ell_2$, the balance is not obvious. 

As discussed in Section \ref{sec:fluc}, the $20$ real scalar fields in ${\bf 10}_c$ of $SU(4)$ branch into ${\bf 1}_c$, ${\bf 3}_c$, and ${\bf 6}_c$ tensors of $SU(3)$ in the KNAdS background.
From the ${\cal N}=2$ supergravity point of view, the ${\bf 1}_c$
are part of the universal hypermultiplet, the ${\bf 3}$ are in massive vectors multiplets, and ${\bf 6}_c$ are tensors under $SU(3)$ but hypermultiplets under supersymmetry. The properties from Table \ref{tab:scallabel} give the effective masses:
\begin{itemize}
    \item $\varphi^{(\mathbf{1})}$: \ $m^2  _{\varphi(\mathbf{1})} = -3 + 9\mathcal{A}_\mu \mathcal{A}^\mu $ (charge $\pm 3$, no Pauli term)~, \\ 
    \item $\varphi^{(\mathbf{3})}$: \ $m^2 _{\varphi(\mathbf{3})}= -3 + \mathcal{A}_\mu \mathcal{A}^\mu $ (charge $\pm 1$, no Pauli term) ~, \\ 
    \item $\varphi^{(\mathbf{6})}$: \ $m^2 _{\varphi(\mathbf{6})}= -3 + \mathcal{A}_\mu \mathcal{A}^\mu -2\mathcal{F}_{\mu \nu} \mathcal{F}^{\mu \nu}$ (charge $\pm 1$, Pauli term with $p=+4$)~.
\end{itemize}
As before, it is instructive to examine the one-parameter black hole families corresponding to Kerr-AdS, BPS, and Reissner-Nordstr\"{o}m-AdS. In the following we do that first, and then we consider general parameters. 

In the Kerr limit, there is no distinction between these three fields, and they all satisfy 
\begin{equation}
m^2 \ell_2 ^2 = \tfrac{3(1-\sqrt{1+8a^2})}{8\sqrt{1+8a^2}}\in (-\tfrac{1}{4},0)~.
\label{eqn:mphikerr}
\end{equation}
Therefore, these scalars are stable in the extremal Kerr-AdS background, but they reach the BF-bound in the limit where the black hole has maximal spin $a\to 1$. It is interesting that so many scalars reach the BF-bound in this very special limit. because it suggests an enhanced symmetry associated with the extremal Kerr/CFT correspondence. 

On the BPS line in the plane of extremal black holes, the effective masses for the pseudo-scalars are:
\begin{align}
\label{eq: varphi1 mass BPS}
    m^2 _{\varphi(\textbf{1})} \ell_2 ^2 & = - \frac{6a(1-a)}{(1+5a)^2}  \in (-\tfrac{1}{4},0) ~, \\
\label{eq: varphi3 mass BPS}
    m^2 _{\varphi(\textbf{3})} \ell_2 ^2 & = - \frac{2a(1+3a)}{(1+5a)^2} \in (-\tfrac{2}{9},0) ~,\\
\label{eq: varphi6 mass BPS}
m^2 _{\varphi(\textbf{6})} \ell_2 ^2 & = \frac{2(2+9a-3a^2)}{(1+5a)^2} \in (\tfrac{4}{9},4) 
   ~. 
\end{align}
All three fields are stable in the AdS$_2$ region. The singlet, which couples to the gauge field with a large charge $e=3$, is exactly at the BF bound on the point on the BPS line where $a=\tfrac{1}{7}$ and $r_+=\tfrac{\sqrt{15}}{7}$. This corresponds to physical charges $Q=\tfrac{7}{18}$ and $J=\tfrac{11}{54}$ (in units of $\tfrac{\pi}{4G_5}$). It is obviously important that $m^2\ell^2_2\geq -\frac{1}{4}$ for all parameters, but there is no clear significance to the bound being saturated at precisely these values. 
As comparison, for BPS black holes, the small black hole branch is the range $0\leq a_{\rm{cusp}}\leq .109$ and the Hawking-Page transition is at $a_{\rm HP}\simeq 0.186$ \cite{Ezroura:2021vrt}.

\begin{figure*}
    \centering
    \begin{subfigure}[t]{0.45\textwidth}
        \centering
        \includegraphics[scale=0.5]{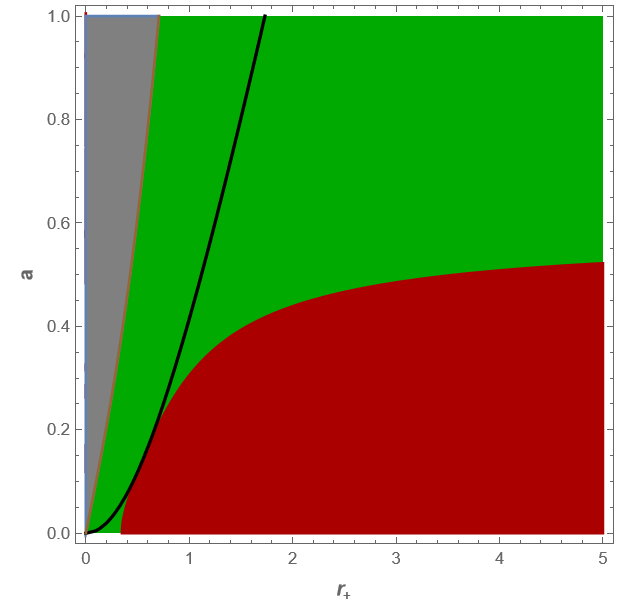}
        \caption{The parameter space for $\varphi^{(\textbf{1})}$. Large minimal couplings drive instability of superconducting type when the extremal black hole is dominated by charge (rather than angular momentum). The unstable region reaches the BPS line at one point.}
 \label{fig:phi1parm1}
 \end{subfigure}\hfill
    ~
    \begin{subfigure}[t]{0.45\textwidth}
        \centering
        \includegraphics[scale=0.5]{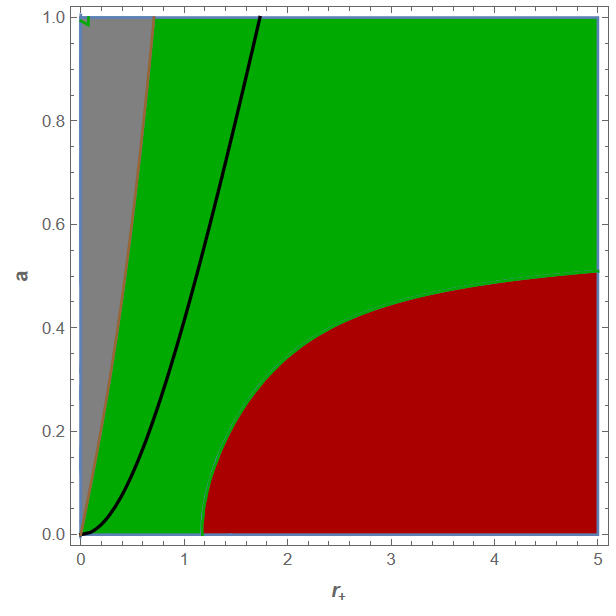}
        \caption{The parameter space for $\varphi^{(\textbf{3})}$. The physics is similar to $\varphi^{(\textbf{1})}$, but the minimal coupling is smaller.}
         \label{fig:phi1parm3}
    \end{subfigure}
    \begin{subfigure}[b]{0.5\textwidth}
        \centering
        \includegraphics[scale=0.5]{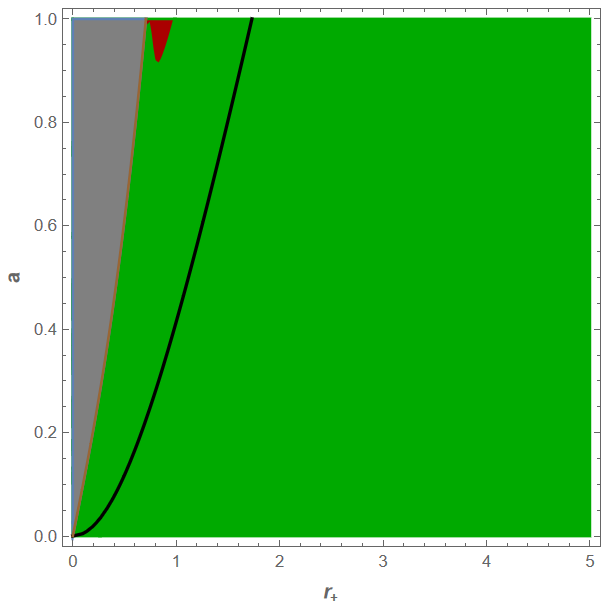}
        \caption{The parameter space for $\varphi^{(\textbf{6})}$. The tiny region of instability is driven by ``magnetic" ${\cal F}_{\mu\nu}{\cal F}^{\mu\nu}>0$ due to the interplay between the rotation and the Chern-Simons terms.}
         \label{fig:phi1parm6}
    \end{subfigure}
    \caption{The parameter space $(r_+,a)$ for the pseudo-scalars $\varphi$. In all three plots, the horizontal axis corresponds to Reissner-Nordstr\"{o}m black holes (no rotation). The grey regions near the vertical axis are unphysical. Their inside boundary (brown line) are the Kerr black hole (no charge). The black line through the middle is the BPS locus \eqref{eqn:rstardef}. Each plot indicates stability $m^2_{\varphi} \ell_2 ^2 >-\tfrac{1}{4}$ (green) and instability $m_{\varphi}^2 \ell_2 ^2 <-\tfrac{1}{4}$ (dark red)}
\end{figure*}

For the Reissner-Nordstr\"{o}m-AdS black holes: 
\begin{align}
 m^2 _{\varphi(\textbf{1})} \ell_2 ^2 & = \frac{-3r_+ ^2 (4+9r_+ ^2)}{4(1+3r_+ ^2)^2}  \in (-\tfrac{3}{4},0) ~,
 \cr
 \ m^2 _{\varphi(\textbf{3})} \ell_2 ^2 & = \frac{-r_+ ^2 (4+11r_+ ^2)}{4(1+3r_+ ^2)^2}\in (-\tfrac{11}{36},0)~, \cr
m^2 _{\varphi(\textbf{6})} \ell_2 ^2 & = \frac{16+76r_+ ^2 +85r_+ ^2}{4(1+3r_+ ^2)^2} \in (\tfrac{85}{36},4) ~.
\end{align}
Because of the large minimal coupling $e=3$, the $\varphi^{(\textbf{1})}$ is driven unstable for nearly all parameters, except for very small RN-black holes. 
The $\varphi^{(\textbf{3})}$ is qualitatively similar but, since $e=1$, it remains stable until a larger value of $r_+$. The $\varphi_{(\textbf{6})}$ is stable for all $r_+$.

There is an additional one-parameter family of black holes that warrants special study: the extremal black holes with $a=1$. On our plots, it corresponds to the {\it upper} horizontal axis. The left end of this line corresponds to neutral black holes that are maximally rotating and, as we noted after \eqref{eqn:mphikerr}, then all pseudoscalars are exactly at the BF-bound. It is interesting to inquire whether they become unstable as we turn on electric charge of the black hole, and so move to the right along the upper horizontal axis. 
One might think so, because, with our sign conventions, an electric field has ${\cal A}_\mu {\cal A}^\mu<0$, and so the contribution of minimal couplings to the effective mass \eqref{eqn:effmass} is destabilizing. This is the intuition that is familiar from holographic superconductors. However, according to Figure \ref{fig:A2 sign}, for extremal black holes with maximal spin $a=1$, we have ${\cal A}_\mu {\cal A}^\mu>0$ for any value of the black hole charge so, when the black hole is maximally rotating, the Lorentz invariant potential ${\cal A}_\mu {\cal A}^\mu$ is magnetic in the near horizon region, even though it is electric asymptotically. Therefore, the minimal coupling to $\mathcal{A}_\mu$ has a {\it stabilizing} effect in the near horizon region of an extremal $a=1$ AdS black hole. Because of this, $\varphi^{(\mathbf{1})}$ and $\varphi^{(\mathbf{3})}$ are stable for $a=1$. That is clear on 
Figures \ref{fig:phi1parm1} and \ref{fig:phi1parm3}.

This is not the entire story, because $\varphi^{(\mathbf{6})}$ also has a coupling directly to the field strength ${\cal F}_{\mu\nu}$, rather than the gauge field ${\cal A}_\mu$. The sign of this ``Pauli" coupling $p=2>0$ is such that, when the field is electric ${\cal F}_{\mu\nu}{\cal F}^{\mu\nu}<0$, the contribution to the effective mass \eqref{eqn:effmass} is positive. Figure \ref{fig:F2 sign} shows that ${\cal F}_{\mu\nu}{\cal F}^{\mu\nu}<0$ for nearly all parameters, and then this contribution is stabilizing as well. However, ${\cal F}_{\mu\nu}{\cal F}^{\mu\nu}$ changes sign in a small sliver near the Kerr limit and, in that tiny region, the Pauli coupling is destabilizing. Moreover, it turns out that the Pauli coupling can dominate the stabilizing ${\cal A}_\mu {\cal A}^\mu$. Therefore, as we consider the $a=1$ black holes 
along the upper horizontal axis, the mode that is at the BF-bound for pure Kerr, turns unstable when the black hole becomes charged. This is the origin of the tiny island of instability in the upper left corner of Figure \ref{fig:phi1parm6}.

\subsection{The Neutral Vectors and their Scalar Mixing} 

Vectors and tensors are more complicated than scalars, for several reasons. Obviously, they have multiple components, and they have gauge symmetry. Therefore, we must deal with polarizations and gauge fixing. However, in addition, none of the $15$ vectors and $12$ tensors fields in gauged ${\cal N}=8$ supergravity are ``minimal", they all have couplings beyond those of a Maxwell field in a curved background $d\star f=0$. On the other hand, we are primarily interested in the AdS$_2$ region, and there we expect that all physical fields ultimately become equivalent to scalar fields, although with non-trivial effective masses due to all the applicable couplings. Our goal is to compute these effective masses in the near horizon region and determine if they correspond to stable modes. 

Among the $27$ original vector fields in ${\cal N}=8$ ungauged supergravity, we do not study ``the" vector field that the EMAdS black hole background is charged with respect to. The remaining $26$ fields are summarized in Table \ref{tab:veclabel}. The $12$ fields that were dualized to tensors are not minimal, they are charged with respect to the background gauge field ${\cal A}_\mu$. The $14=15-1$ ordinary gauge fields that vanish in the background include $6$ components $a^+$ that are not only charged with respect to the background gauge field ${\cal A}_\mu$, they also couple directly to the background field strength ${\cal F}_{\mu\nu}$. The final $8$ vector fields $a_-$ are simpler than $a_+$. because they are neutral with respect to the background gauge field, but, they couple to the $8$ scalar fields $t_-$. 
In this subsection we address the neutral fields described by the coupled system $(t_-, a_-)$. 

The equations of motion given in 
(\ref{eqn:tm eom original}-\ref{eqn:am eom original})
can be presented as $8$ decoupled blocks, each of the form: 
\begin{align} \label{eq: tm eom} 
        d \star d t_- + 4g^2 t_-  \star 1 -t_- ({\cal F}\wedge \star {\cal F}) &=  2{\cal F} \wedge \star f_- ~, \\ 
        \label{eq: fm eom}
d \star  f _- - {\cal F} \wedge  f _- &= 2d \left( t_- \star \mathcal{F}\right) ~.
\end{align}
The one-form $f_-=da_-$ describes the fluctuating vector field, while ${\cal F}$ refers to the field strength of the background black hole. 

In the near horizon region it is sufficient to consider fields that depend on the $(t,r)$ coordinates within $\text{AdS}_2$. Accordingly, we decompose the one-form $a_-$ into five component fields $\{a^-_t, a^-_r, a^-_i\}$ on AdS$_2$ as:
\begin{equation}
\label{eq: ap decomp}
    a_- = a_t ^- dt + a_r ^- dr + \sum_{i=1}^3 a_i ^- \sigma_i ~.
\end{equation}
For a gauge field in five dimensions, we expect that gauge symmetry removes one of the five component fields and a constraint removes another, leaving three propagating degrees of freedom. The situation is less clear in two dimensions but it turns out that, from our point of view, the intuition from five dimensions offers good guidance. To proceed, we insert the general expansion \eqref{eq: ap decomp} into the equations of motion \eqref{eq: fm eom}, without picking a gauge. This gives the vanishing of a complicated four-form or, equivalently, $5$ differential equations. The scalar equation \eqref{eq: tm eom} gives yet another, for a total of $6$. 

The two differential equations that correspond to the preserved $S^2$, and so the labels $1, 2$, are the simplest. That is  because they only involve the 2D scalar fields $a^-_1, a^-_2$ that have polarizations along these directions. In the near horizon region they become: 
\begin{align}
    \left[\frac{1}{A^2 (r-r_+)^2} \partial_t ^2 -\frac{1}{B^2}\partial_r ((r-r_+)^2 \partial_r) + \left(\frac{4}{D^2} - \frac{E^2}{A^2} -\frac{2F}{ABD}\right)\right] a_1 ^- &= \frac{2E}{A^2 (r-r_+)} \partial_t a_2 ^-  ~, \\ 
    \left[\frac{1}{A^2 (r-r_+)^2} \partial_t ^2 -\frac{1}{B^2}\partial_r ((r-r_+)^2 \partial_r) + \left(\frac{4}{D^2} - \frac{E^2}{A^2} -\frac{2F}{ABD}\right)\right] a_2 ^- &= -\frac{2E}{A^2 (r-r_+)} \partial_t a_1 ^-  ~.
\end{align}
The two fields $a_1$ and $a_2$ {\it nearly} decouple from one another, and they decouple completely in the complex basis: 
\begin{equation} \label{eq: ap perp def}
    a_\perp ^-= a_1 ^- +i a_2 ^- ~,
\end{equation}
where
\begin{equation} \label{eq: am perp eom}
    \left[\frac{1}{A^2 (r-r_+)^2} \partial_t ^2-\frac{1}{B^2}\partial_r ((r-r_+)^2 \partial_r) + \left(\frac{4}{D^2} - \frac{E^2}{A^2} -\frac{2F}{ABD}\right)\right] a_\perp ^- = -\frac{2iE}{A^2 (r-r_+)} \partial_t a_\perp  ^- ~.  
\end{equation}
Because of the first-order time derivative term on the right hand side, this equation of motion for $a^-_\perp$ is not exactly the same as the canonical free massive scalar equation: 
\begin{equation}
    \label{eq: canonical free massive kg eom}
    \left(\Box_2 - m^2_{\rm eff}\right) \chi = 0 ~,
\end{equation}
for some generic scalar $\chi$ with mass $m_{\rm eff} $, and where $\Box_2$ is the near-horizon $\text{AdS}_2$ kinetic operator:
\begin{equation}
\label{eq:Box AdS2 def}
\Box_2 = -\frac{1}{A^2 (r-r_+)^2} \partial_t ^2 +\frac{1}{B^2}\partial_r ((r-r_+)^2 \partial_r) ~.
\end{equation}
However, for the purposes of analyzing stability, the singular potential $\tfrac{1}{r-r_+}$ is subdominant for $r\to r_+$, and the normalizability condition at the heart of the BF argument is unaffected by that additional term. With this understanding, we read off the effective mass for the transverse polarizations: 
\begin{equation}
m^2_{\perp-}
=\frac{4}{D^2} -\frac{E^2}{A^2} -\frac{2F}{ABD}~.
\label{eqn:mefftrans}
\end{equation}

The remaining four differential equations couple the fields $(a_t ^-, a_r^-, a_3 ^-, t_-)$. However, the two components in the AdS$_2$ directions are nearly trivial: 
\begin{align}
0  & =  d \left( \partial_r a^-_t - \partial_t a^-_r   -Er\partial_r a^-_3  + 2F t_- + \frac{8ABG}{DC^2} a^-_3\right)  ~.
\label{eqn:totd}
\end{align}
This is solved by 
\begin{equation}
    \partial_r a^-_t - \partial_t a^-_r   -Er\partial_r a^-_3  + 2F t_- + \frac{8ABG}{DC^2} a^-_3=0~, 
    \label{eqn:gaugecond}
\end{equation}
which serves as both gauge condition and the dual Gauss' law constraint. 
A nonvanishing constant on the right hand side of \eqref{eqn:gaugecond} would represent a constant background charge density for the fluctuating field. Such a global feature may be interesting, but it is not relevant in our context. 

The equations of motion that were not yet analyzed are the scalar equation \eqref{eq: tm eom} and the component of the vector equation \eqref{eq: fm eom} with polarization along the $3$ direction, corresponding to the direction of black hole rotation. After imposing the gauge condition \eqref{eqn:gaugecond}, and some nontrivial algebra, these equations become equivalent to two coupled scalar fields:  
\begin{equation}
    \Box_2 \begin{pmatrix}
        a_3 ^- \\ 
        t^-
    \end{pmatrix} - M^2_{\rm eff} \begin{pmatrix}
        a_3 ^- \\ 
        t^-
    \end{pmatrix} = 0 ~, 
\end{equation}
where the $2\times 2$ mixing matrix: 
\begin{equation}
M^2_{\rm eff} = \begin{pmatrix}
        \frac{4D^2}{C^4} -\frac{2D(F-EG)}{ABC^2}+\frac{16G^2}{C^4} & \frac{16DG}{C^4} +\frac{8FG}{ABC^2}  \\ 
       \frac{16DG}{C^4} + \frac{8FG}{ABC^2}   & -4g^2 +\frac{16G^2}{C^4}+\frac{3F^2}{A^2 B^2} 
    \end{pmatrix} ~.
    \label{eqn:mixingmass}
\end{equation}
The contribution $4D^2/C^4$ to the mass of the vector is analogous to $4/D^2$ for the other polarizations \eqref{eqn:mefftrans}. It is positive and due to the minimal orbital angular momentum required for a vector field. 

For $t_-$, the $-4g^2$ is the AdS$_5$ mass. Because the Pauli coupling is negative $p=-2$ for $t_-$, the $F_{\mu\nu}F^{\mu\nu}$ terms gives an electric contribution $-\frac{F^2}{A^2 B^2}$ that also drives instability. However, it turns out that the coupling to the gauge field $f_-$ on the right hand side of \eqref{eq: tm eom}, recast using the gauge condition \eqref{eqn:gaugecond}, gives a term of the same form that compensates, and then some, for a total of $+\frac{3F^2}{A^2 B^2}$.

To summarize so far: we have analyzed the coupled equations of motion (\ref{eq: tm eom}-\ref{eq: fm eom}) for the fields $(t_-,a_-)$. There are four physical degrees of freedom, as expected, because a vector has three polarizations in five dimensions and a scalar has one. Their effective masses are \eqref{eqn:mefftrans} (with multiplicity $2$) and the eigenvalues of the mixing matrix \eqref{eqn:mixingmass}. In the following, we study the BF-stability bound 
$m_\text{eff}^2 \ell_2 ^2\geq - \frac{1}{4}$ based on these formulae. 

\begin{figure}[h]
    \centering    \includegraphics[width=0.5\linewidth]{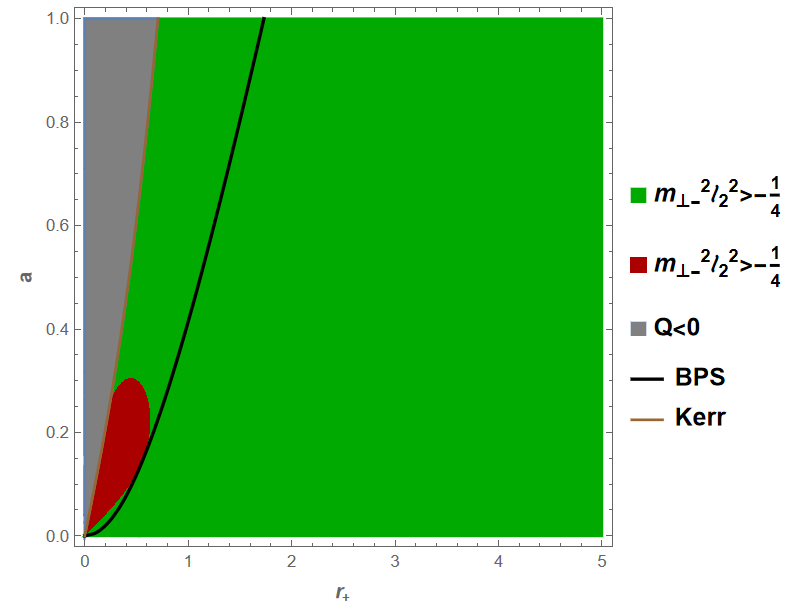}
    \caption{BH stability parameter space for the transverse fields $a_{1,2} ^-$, where $m_{\perp-} ^2 \ell_2 ^2 \geq -\tfrac{1}{4}$ (green) and $m_{\perp-} ^2 \ell_2 ^2 \leq -\tfrac{1}{4}$ (dark red), and the unphysical region $Q<0$ (grey). Fluctuations in $a_\perp ^-$ are stable except for a finite region of small, slowly rotating black holes.}
\label{fig:am perp stab}
\end{figure}

As in previous cases, it is instructive to first consider the one parameter black hole families corresponding to Kerr-AdS, BPS, and Reissner-Nordstr\"om-AdS. For the transverse fields $a^-_{1,2}$ with polarizations along the $S^2$ and effective mass \eqref{eqn:mefftrans}, we have
\begin{align}
\label{eq: aperp min mass Kerr}
    &\text{Kerr: } ~~~m_{\perp-}^2 \ell_2 ^2 =- \frac{8\left(1-a^2\right)^2}{(8a^2+1)\left(3 +\sqrt{8
   a^2+1}\right)^2}  \in (-\tfrac{1}{2},0)  ~, \\
\label{eq: aperp min mass BPS}
    &\text{BPS: } ~~~~m_{\perp-}^2 \ell_2 ^2 = -\frac{6a(1-a)}{(1+5a)^2} \in (-\tfrac{1}{4},0)~, \\ 
\label{eq: aperp min mass RN}
    &\text{RN:} ~~~~~~~m_{\perp-}^2 \ell_2 ^2 = \frac{1-\sqrt{1+2r_+ ^2}}{1+3r_+ ^2} \in (\sqrt{\tfrac{2}{3}}-1,0) ~.
\end{align}
One the Kerr line, the effective mass for $a_{1,2} ^-$ reaches below the BF bound for the range $0\leq a \leq \tfrac{1}{2}\sqrt{13-9\sqrt{2}})\simeq 0.261$. On the BPS line, the effective mass for $a^-_{1,2}$ is identical to the result for the pseudoscalar $\varphi(\textbf{1})$ \eqref{eq: varphi1 mass BPS}, including the feature that the BF-bound is reached for exactly one, non-trivial, value of the rotation parameter $a=\frac{1}{7}$. As comparison, for BPS black holes, the small black hole branch is the range $0\leq a_{\rm{ cusp}}\leq .109$ and the Hawking-Page transition is at $a_{\rm HP}\simeq 0.186$ \cite{Ezroura:2021vrt}. Figure \ref{fig:am perp stab} plots the unstable region on the entire parameter space.


\begin{figure}[h]
    \centering    \includegraphics[width=0.5\linewidth]{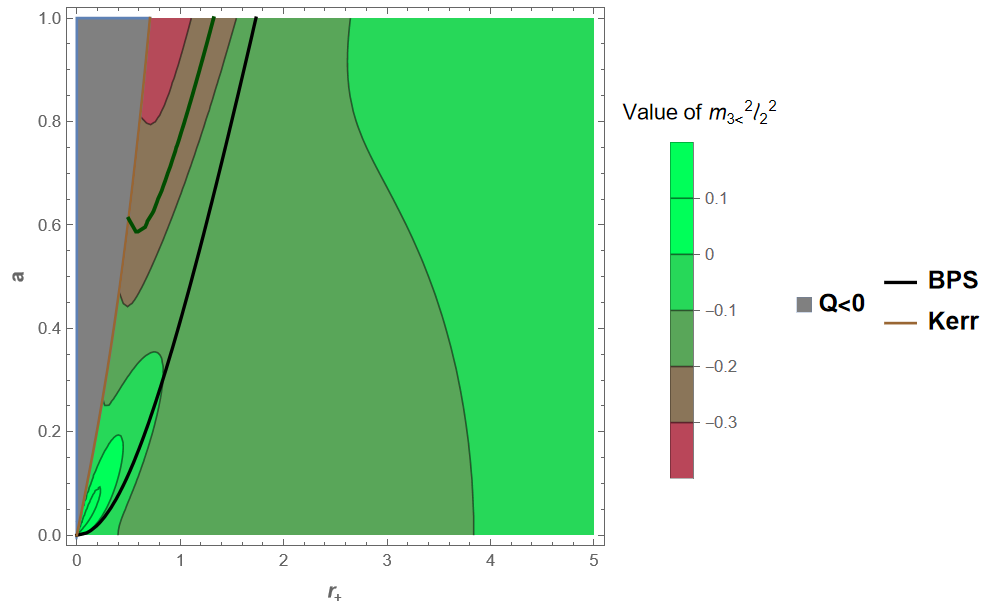}
    \caption{BH stability parameter space for the lower-mass $(a_3 ^-, t_-)$ diagonalized field, in the form of a contour plot for the value of $m_{3<} ^2 \ell_2 ^2$ and the unphysical region $Q<0$ (grey). The $m_{3<} ^2 \ell_2 ^2 = -\tfrac{1}{4}$ boundary is in dark green, the BPS line in brown and the Kerr line in green. Apart from a finite region of fast-spinning black holes, the lower-mass diagonalized field is stable.}
\label{fig:a3mstab}
\end{figure}

For the physical $a_3$ and $t_-$, and their mixing, we study the mixing matrix \eqref{eqn:mixingmass}. Generally, the eigenvalues of this matrix are complicated functions of $(A,B,C,D,E,F,G)$, but for the benchmark curves in parameter space (Kerr, BPS, Reissner-Nordstr\"om) the analytical formulae are manageable:
\begin{align}
    \label{eq: a3 tm mix mass Kerr}
    &\text{Kerr: } \{m_{3<} ^2,m_{3>} ^2\} \ell_2 ^2 =  \left\{\tfrac{\sqrt{1+8a^2}-1}{2\sqrt{1+8a^2}}, \tfrac{2}{\sqrt{1+8a^2}}\right\} \in \{(-\tfrac{1}{3},0),(\tfrac{2}{3},2)\}~, \\
    \label{eq: a3 tm mix mass BPS}
    &\text{BPS: } \{m_{3<} ^2,m_{3>} ^2\} \ell_2 ^2 =\left\{\frac{3+2a\pm \sqrt{9+20a+20a^2}}{2(1+5a)}\right\} \in \{(-\tfrac{1}{6},0),(1,3)\} ~, \\ 
    \label{eq: a3 tm mix mass RN}
    &\text{RN: } \{m_{3<} ^2,m_{3>} ^2\} \ell_2 ^2 = \left\{\tfrac{4+5r_+ ^2 +\sqrt{1+2r_+ ^2}\pm \sqrt{5+22r_+ ^2 +25r_+ ^4 +2(1+5r_+ ^2)\sqrt{1+2r_+ ^2}}}{2(1+3r_+ ^2)}\right\} \cr
    &\qquad \qquad \qquad \qquad \in \{(-1+\sqrt{\tfrac{2}{3}},0),(\tfrac{5}{3},3)\}~.
\end{align}
The higher-mass field, the one with effective mass $m_{3>}$, is stable not only along each of these lines, but throughout the physical parameter space. The lower mass field has effective mass $m_{3<}$ and is stable by the BF-criterion along the BPS and RN lines, On the Kerr line $a_{3<} ^-$ is only stable in the range $0<a<\tfrac{\sqrt{3}}{2\sqrt{2}}$. As it happens, for this mode the effective mass on the Kerr-line is that same as the one for $m^2_{t_+}$. In Figure \ref{fig:a3mstab} we plot the physical parameter space of $a_{3<} ^-$ as a contour plot. This variation over previous plots highlights that we did compute the effective mass throughout parameter space, even though have focused on the issue of stability. 

\subsection{The Charged Vectors and Tensors} 
The vector fields $a_-$ addressed in the previous subsection experience a Pauli-type coupling directly to the background field strength ${\cal F}$, and also mixing with a scalar field, but they are neutral with respect to the background gauge field ${\cal A}$.  
The remaining vector fields $a^+$ satisfy equations of motion of the form 
\begin{equation}
D \star  f^+ + {\cal F}\wedge f^+ =0~, 
\label{eqn:fp schem}
\end{equation}
where $f^+=Da^+$. When comparing with its analogue for $a^-$ \eqref{eq: fm eom}, the Pauli-coupling to ${\cal F}$ has the opposite sign, and there is no scalar field. However, the most important difference is 
the minimal coupling encoded in the gauge covariant derivative $D = d + ie{\cal A}$ with $e=2$. The $i$ appears when diagonalizing the equations of motion \eqref{eqn:ap eom}, by taking linear combinations that account for the $\Omega$-operation. 
There are three independent vector fields that satisfy \eqref{eqn:fp schem}, and three others satisfy its complex conjugate. The $\mathbf{6}_c$ two-form tensor fields $B_{\mu \nu} ^{i\alpha}$ similarly satisfy \eqref{eq: Bfield eom}
\begin{equation}
    D B  =  i \star B~,
    \label{eqn:Bfield eomform}
\end{equation}
where, according to \eqref{eq: Bfield charge}, the gauge covariant derivative $D = d + i e{\cal A}$ now has $e=1$.


In all these equations, the charged vectors and tensors enjoy two kinds of gauge invariance. Their transformation as matter fields $a^+\to e^{-2i\Lambda}a^+$, $B\to e^{-i\Lambda}B$ is a symmetry when supplemented by transformation of the background vector field as ${\cal A}\to {\cal A}+d\Lambda$. This gauge symmetry is fixed by the generalized Lorentz condition $d\star {\cal A}=0$ satisfied by our background gauge field. The second symmetry generalizes $a^+ \to a^+ + d\lambda$ to take the background gauge field into account. It is responsible for removing local degrees of freedom, as usual for a gauge symmetry. Unfortunately, it is a great deal more complicated, due to the minimal coupling. This mechanism must work, because of the symmetries involved, but we have not carried out an explicit construction. 

We can make some progress by proceeding as in previous cases. As in \eqref{eq: ap decomp}, we decompose the vector field in components: 
\begin{equation}
\label{eq: app decomp}
    a_+ = a_t ^+ dt + a_r ^+ dr + \sum_{i=1}^3 a_i ^+ \sigma_i ~. 
\end{equation}
Inserting this expansion into \eqref{eqn:fp schem}, we find five differential equations. The two that correspond to polarizations $1$ and $2$, along the $S^2$, decouple from the others and are relatively simple. 
In analogy with \eqref{eq: ap perp def}, we introduce the complex field:
\begin{equation} \label{eq: app perp def}
    a_\perp ^+= a_1 ^+ +i a_2 ^+ ~.
\end{equation}
Then these two polarizations satisfy: 
\begin{align} 
    &\left[ \frac{1}{A^2 (r-r_+)^2} \Big(\partial_t + 2i E(r-r_+)\Big)\partial_t-\frac{1}{B^2}\partial_r ((r-r_+)^2 \partial_r) + m^2_\perp\right]a_\perp ^+ 
    =0 ~, 
\end{align}
and its complex conjugate. This is analogous to \eqref{eq: am perp eom}. The effective 2D mass becomes
\begin{equation}
m^2_{\perp+} = \frac{4}{D^2} + \frac{2F}{ABD} - \frac{E^2}{A^2} +\frac{16G^2}{D^2} - \frac{4(F-EG)^2}{A^2}~.
\label{eqn:atransplus}
\end{equation}
Comparing with the effective mass of $a^-$ \eqref{eqn:mefftrans}, the first three terms are the same, except that the Pauli term $\frac{2F}{ABD}$ has the opposite sign. The remaining terms are due to the mass type contribution $e^2 \mathcal{A}_\mu \mathcal{A}^\mu$ with $e=2$. It is interesting that the $\frac{16G^2}{D^2}$ term is reminiscent of the effective mass for $a^-_3$.

\begin{figure}[h]
    \centering    \includegraphics[width=0.5\linewidth]{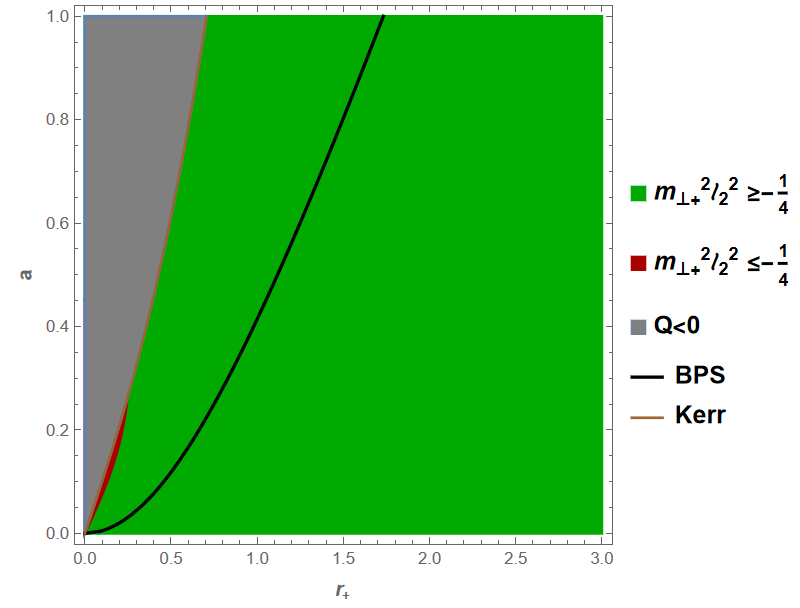}
    \caption{Black hole instability region for the $a^+_{1,2}$ fields, where $m_{\perp+} ^2 \ell_2 ^2 \geq -\tfrac{1}{4}$ (green) and $m_{\perp+} ^2 \ell_2 ^2 \leq -\tfrac{1}{4}$ (dark red), and the unphysical region $Q<0$ (grey). $a_{1,2}$ is stable for nearly the entire physical domain, except for a small sliver close to the Kerr line.}
    \label{fig:aperpstab}
\end{figure}

Generally, fields with spin are stabilized by their angular momentum which cannot vanish. However, in a rotating background angular momentum is not conserved, and there is a complicated interplay between geometric and electromagnetic contributions. As a probe we consider the transverse fields $a_{1,2} ^+$ with the exact effective mass \eqref{eqn:atransplus}. In AdS$_2$, vector fields satisfy the same BF bound as a scalar $m_{\perp+} ^2 \ell_2 ^2 \geq -\frac{1}{4}$. For the three representative lines through parameter space: 
\begin{align}
\label{eq: aperp plus mass Kerr}
    &\text{Kerr:}~~ m^2_{\perp+} \ell_2 ^2 = 
    - \frac{8\left(1-a^2\right)^2}{(8a^2+1)\left(3 +\sqrt{8
   a^2+1}\right)^2} \in (-\tfrac{1}{2},0)  ~, \\
\label{eq: aperp plus mass BPS}
    &\text{BPS:}~~ m_{\perp+} ^2 \ell_2 ^2 =\frac{2-10a^2}{(1+5a)^2} \in (-\tfrac{2}{9},2)~, \\ 
\label{eq: aperp plus mass RN}
    &\text{RN:}~~~ m_{\perp+} ^2 \ell_2 ^2 = \frac{1+2r_+ ^2 -2r_+ ^4  + (1+3r_+ ^2)\sqrt{1+2r_+ ^2}}{(1+3r_+ ^2)^2} \in (-\tfrac{2}{9},2) ~.
\end{align}
This shows the vector modes $a_\perp ^+$ are stable along the entire Reissner-Nordstr\"{o}m and BPS lines, as expected by the heuristic argument. However, on the Kerr line, there is an instability for small rotation. The plot in figure \ref{fig:aperpstab} confirms that the mode is {\it nearly} always stable, but there is an exception at a sliver near slowly rotating Kerr. We have no intuition for why that is the case. 

We still did not consider the longitudinal polarization. The two differential equations that include $\sigma_1\sigma_2\sigma_3$ in four-form language are ``along" AdS$_2$. Before taking the $U(1)$ charge into account these two are a perfect gradient on AdS$_2$, which can be integrated. Then one equation can be interpreted as the gauge condition, and the other as a constraint, as in \eqref{eqn:totd}. In the case here, the closest generalization we have found is the gauge condition 
\begin{equation}
\partial_r a^+_t- \partial_t a^+_r 
- E(r-r_+)\partial_r a_3 ^+ + 2(F-EG)(r-r_+)ra_r ^+ - \frac{8ABG}{C^2D} a_3 ^+=0 ~,
\label{eqn:gengaugc}
\end{equation}
on the AdS$_2$ base space. This solves both AdS$_2$ equations exactly, when the magnetic field vanishes ($G=0$), but generally there are additional nonlinearities that we can only account for perturbatively. 

The fifth differential equation, the one that is along the $S^1$ fibre, is by far the most complicated. However, with the gauge condition \eqref{eqn:gengaugc}, and the caveat above, it reduces to a standard scalar equation for the field $a^+_3$. It has effective mass
\begin{equation}
m^2_{3+} = \frac{4D^2}{C^4} + \frac{2D(F-EG)}{ABC^2} - \frac{4(F-EG)^2}{A^2}+ \frac{16G^2}{C^4}~.
\label{eqn:alongplus}
\end{equation}
The contributions involving $G$ are heuristic, in the sense that other terms involving $G$ were not computed.  

Evaluating \eqref{eqn:alongplus} along the three representative lines in parameter space, we find: 
\begin{align}
    \label{eq: a3 p mass Kerr}
    &\text{Kerr:}~~ m^2_{3+} \ell_2 ^2 = 
   \tfrac{2}{\sqrt{1+8a^2}}\in (\tfrac{2}{3},2)  ~, \\
    \label{eq: a3 p mass BPS}
    &\text{BPS:}~~ m_{3+} ^2 \ell_2 ^2 =\frac{4+24a+27a^2-a^3}{(2+a)(1+5a)^2} \in (\tfrac{1}{2},2)~, \\ 
    \label{eq: a3 p mass RN}
    &\text{RN:}~~~ m_{3+} ^2 \ell_2 ^2 = \frac{1-2r_+ ^4 +\sqrt{1+2r_+ ^2}+r_+ ^2 (2+3\sqrt{1+2r_+ ^2})}{1+3r_+ ^2} \in (-\tfrac{2}{9},2) ~.
\end{align}
These formulae suggest that the $a_3 ^+$ field is stable on all three curves. Indeed, upon further examination, $m_{3+} ^2 \ell_2 ^2$ is greater than $-\tfrac{1}{4}$ for all physical $(r_+ , a)$, which implies stability throughout parameter space. As argued previously, we do indeed expect fields with spin to be notably stable, because they do not allow for an s-wave. 


The two-form tensor $B$ has three degrees of freedom, after all gauge conditions have been fixed. They can be parametrized by the three pairs of indices on $S^3$. Then there are two degenerate polarizations, the ones with indices $31$ and $23$, while the third $12$ is distinct. The three polarizations have a common contribution: 
\begin{align}
         m^2 _B &= 1+ \mathcal{A}^\mu \mathcal{A}_\mu = 1 + \frac{4G^2}{D^2} - \frac{(F-EG)^2}{A^2}
~.
\label{eqn:MBresult}
\end{align}
This is analogous to a scalar with the coupling $e=1$. 
Since $\mathcal{A}_\mu \mathcal{A}^\mu$ \eqref{eqn:Asqdef} is bounded from below by $-1$ on the entire parameter space, this effective mass is non-negative. Unfortunately, there is an additional contribution that we have been unable to gauge fix properly. It serves to split the three modes. This omission is unlikely to change the stability, since this field has spin. 


\subsection{Summary} 
\label{subsec:fluceffmass}

In this section we studied all bosonic matter fields of ${\cal N}=8$ gauged supergravity, in the AdS$_2$ near horizon  region of the extremal Kerr-Newman AdS black holes. We are able to express (nearly) all modes in terms of an equivalent scalar field, so we can present the spectrum in terms of the effective masses of these scalars. Each mass is equivalent to a conformal dimension in the dual CFT$_1$, but we do not use that terminology. 

In this paper, we generally follow the standard  AdS$_5$ convention that the  radius $\ell_5=g^{-1}=1$. However, in this section, the most important scale is the AdS$_2$ radius $\ell_2=B^{-1}$ which varies greatly over the black hole parameter space. We address this by restoring the coupling constant of gauged supergravity $g$, when it is needed for emphasis. 

The spectrum depends on the two black hole parameters, the charge and the angular momentum, in a complicated way. We find general parametric formulae but, to understand their qualitative significance, we present partial results in two complementary ways.

\begin{table} [h]
    \centering
    \begin{tabular}{c|c|c|c|c|c|c|c|c|c|c}
        \text{~Field~} & $~~m_{\text{AdS}_5~~} ^2$ & $~\Delta_{\text{AdS}_5}~$ & $~~~e~~~$ & $~~~p~~~$ & Effective mass $m_\text{eff} ^2$ & $m_\text{BPS} ^2 \ell_2 ^2 $  & ~\#~ \\
        \hline
         $t^+$ & $-4g^2$ & 2 & $\pm 2$ & $+2$ & $-4g^2 +4g^2 \mathcal{A}_\mu \mathcal{A}^\mu -\frac{1}{2} \mathcal{F}_{\mu \nu} \mathcal{F}^{\mu \nu} $ & \eqref{eq: tplus mass BPS} & 12 \\
         \hline 
         $t^-$& $-4g^2$ & 2 & 0 & $-2$ & $m_{t^-} ^2 $ (mixed) & \eqref{eq: a3 tm mix mass BPS}  & 8 \\
         \hline
         $\Lambda^\alpha _\beta$ & 0 & 4 & 0 & 0 & 0 & 0 & 2\\
         \hline
         $\varphi_{ijk\alpha} ^\mathbf{1}$& $-3g^2$ & 3 & $\pm 3$ &  0 & $-3g^2 + 
         9g^2\mathcal{A}_\mu \mathcal{A}^\mu $ & \eqref{eq: varphi1 mass BPS}
         &  2 \\
         \hline
         $\varphi_{ijk\alpha} ^\mathbf{3}$& $-3g^2$ & 3  &  $\pm 1$ &  0 & $-3g^2 + g^2\mathcal{A}_\mu \mathcal{A}^\mu $ & \eqref{eq: varphi3 mass BPS}&  6\\
         \hline
         $\varphi_{ijk\alpha} ^\mathbf{6}$& $-3g^2$  & 3 & $\pm 1$ &  0 & $-3g^2 + g^2\mathcal{A}_\mu \mathcal{A}^\mu $ & \eqref{eq: varphi6 mass BPS}
          &  12  \\
         \hline
    \end{tabular}
    \caption{Table of effective masses effective near horizon for the $42$ scalar fluctuations.}
    \label{tab:scal flucts}
\end{table}

We compute formulae for the effective masses on three important curves through parameter space: Kerr-AdS, BPS, and Reissner-Nordstr\"{o}m-AdS. For BPS and Kerr, it is convenient to parametrize the curve by the rotation parameter $a\in (0,1)$. The Reissner-Nordstr\"{o}m black hole does not rotate so $a=0$, and we use instead the radial coordinate position of the horizon $r_+$. For each mode, the analytical expressions for these three curves give one impression of the effective mass for general parameters. 

\begin{table} [h]
    \centering
    \begin{tabular}{c|c|c|c|c|c|c|c|c}
        ~\text{Field} ~& $~m_{\text{AdS}_5} ^2~$ & $\Delta_{\text{AdS}_5}$ & $~~~e~~~$  & Effective mass $m_\text{eff} ^2$  & $m_\text{BPS}^2 \ell_2 ^2$ & ~~\#~~\\
        \hline
        $a^+ _{1,2}$ & $0$ & 3 & $\pm 2$ & $m_{\perp +} ^2$ &\eqref{eq: aperp plus mass BPS} 
        & $6\times 2$\\ 
        \hline 
        $a^+ _3$ & $0$ & 3 & $\pm 2$ & $m_{3+} ^2 $ & \eqref{eq: a3 p mass BPS} & 6\\
        \hline
        $a^- _{1,2} $ & $0$ & 3 & $0$ & $m_{\perp -} ^2$ & \eqref{eq: aperp min mass BPS}  & $8 \times 2$\\
        \hline
        $a^- _3 $ & $0$ & 3 & $0$ & $m_{a^-} ^2$ (mixed) & \eqref{eq: a3 tm mix mass BPS} & 8\\
        \hline
        $B^{\pm}$ & $+g^2$ & 3 & $\mp 1 $ &  $m^2_B$ &  & $12\times 3$ \\
        \hline
    \end{tabular}
    \caption{Table of effective near horizon masses for the $26$ vector fluctuations. The $\#$ refers to the multiplicity with ``$\times$" referring to the the number of polarizations.}
    \label{tab:vec flucts}
\end{table}

As a complementary representation, for most fields we plot the stability region, as measured by the BF stability bound $m^2_{\rm eff}\ell_2^2\geq -\frac{1}{4}$, for all black hole parameters. It is significant that, for all fields, the BPS curve is in the stable region. However, in some cases it is right at the bound. The details of the non-minimal couplings we find are essential for this outcome. Conversely, this feature give us some confidence in our results.  

The tables in this subsection are complementary to those presented in section \ref{subsec:fluceom}. Here our focus is entirely on the near-horizon region. It is not realistic to include general formulae in the tables, so we highlight in particular the explicit formulae for the BPS masses. 

We give no effective masses for $a_3^+$ and $B^\pm$ because, as discussed in the text, our formulae account only for some of the magnetic effects. In particular, the three $B$ polarizations are split into $2+1$ by effects that we did not compute.

\section{Discussion}
\label{sec:discussion}

The question that motivated this work is whether extremal black holes are stable, both classically and at the quantum level. Basic thermodynamics suggests they are not, but does not shed light on the nature of the instability. Quite sensibly, exploratory research in the area has focused on simple backgrounds, such as Reissner-Nordstr\"{o}m black holes, and simple families of perturbations, like minimally coupled scalar fields with any mass. However, in these settings it is unclear whether the theory is stable at the quantum level, even if no black hole is involved. This drives us to study ${\cal N}=8$ AdS supergravity, which is clearly a well-behaved theory, and some of its relatives. However, the conceptual clarity guaranteed by a clear ground state gives rise to substantial technical challenges. 

To ensure stability, we consider black holes families that include a supersymmetric locus. In AdS, that requires rotation, a significant complication. Moreover, for consistency, we study the matter given by the theory. That turns out to involve unfamiliar non-minimal couplings, another significant complication. These refinements are important. If the non-minimal couplings are ignored, we can establish that extremal black holes are unstable in a sense that is too strong, because it ``shows" that BPS black holes are unstable. When the correct couplings are taken into account, BPS black holes stable, but large swaths of the remaining parameter space is not. 

The specifics are somewhat surprising. For example, the ${\bf 20}'$ modes with $m^2\ell^2_5=-4$ are just at the margin of stability in the AdS$_5$ vacuum, and so one might think that the black hole background drives them unstable. However, it turns out that, when all the non-minimal couplings are taken into account, the ${\bf 10_c}$ modes with $m^2\ell^2_5=-3$ condense over a broader region of parameter space. 

Our results represent progress, but they are not complete. The equations we find apply throughout spacetime, but we only analyze them in the near horizon region. Also, for parameter where we identity specific unstable modes, we do not identify the end-product of the decay. In this sense, we determine the {\it onset} of various instabilities. It would be interesting to follow them to their endpoint. Moreover, it is entirely possible that the fields retained in 5D gauged supergravity are insufficient for a complete description. For example, for some processes the full 10D supergravity, including modes that are localized on the $S^5$, may be needed \cite{Kim:2023sig}. 

One of the most interesting open questions is the nature of the phase transition at the BPS line. The extremal black holes have electric potential $\Phi>1$ all the way from the Reissner-Nordstr\"{o}m line to the BPS curve, and they similarly have $\Omega>1$ all the way from the Kerr line to the BPS curve. This suggests phases that are superconducting and fragmented by superradiance, respectively. In the dual CFT, such phases have natural interpretations in terms of Bose-Einstein condensation, with distinct modes condensing in the two phases. It would be interesting to establish these interpretations firmly.

\section*{Acknowledgements}

We thank Seok Kim and Siyul Lee for discussions and Shiraz Minwalla for correspondence. 
This work was supported in part by DoE grant DE-SC0007859. NE thanks the Leinweber Center for Theoretical Physics for support. FL thanks Stanford ITP and ICTS Bengaluru for hospitality while this article was completed.

\bibliographystyle{JHEP}
\bibliography{main.bib}

\end{document}